\documentclass[aps,twocolumn]{revtex4}  
\usepackage{graphicx}  
\usepackage{epstopdf}
\usepackage{dcolumn}   
\usepackage{bm}        
\usepackage{amsmath}
\usepackage{amssymb}   
\usepackage{array}
\usepackage[caption=false]{subfig}
\usepackage[bookmarks=false,linkcolor=blue,urlcolor=blue,colorlinks,citecolor=blue]{hyperref}
\usepackage{dsfont}
\usepackage{appendix}

\DeclareMathOperator{\Tr}{Tr}
\DeclareMathOperator{\sgn}{sgn}
\makeatletter

\begin{document}


\title{Time-Reversal-Invariant Topological Superconductivity Induced by Repulsive Interactions in Quantum Wires}
\author{Arbel Haim, Anna Keselman, Erez Berg and Yuval Oreg}
\affiliation{Department of Condensed Matter Physics$,$ Weizmann Institute of Science$,$ Rehovot$,$ 76100$,$ Israel}
\date{\today}

\begin{abstract}
We consider a model for a one-dimensional quantum wire 
with Rashba spin-orbit coupling and repulsive interactions, proximity coupled to a conventional {\it s}-wave superconductor. Using a combination of Hartree-Fock and density matrix renormalization group calculations, we show that for sufficiently strong interactions in the wire, a time-reversal invariant topological superconducting phase can be stabilized in the absence of an external magnetic field. This phase supports two zero-energy Majorana bound states at each end, which are protected by time-reversal symmetry. The mechanism for the formation of this phase is a reversal of the sign of the effective pair potential in the wire, due to the repulsive interactions. We calculate the differential conductance into the wire and its dependence on an applied magnetic field using the scattering-matrix formalism. The behavior of the zero-bias anomaly as a function of the field direction can serve as a distinct experimental signature of the topological phase.
\end{abstract}

\maketitle

\emph{Introduction.} Topological phases in condensed matter physics have attracted much attention in recent years~\cite{Qi2011,Hasan2010}. Such phases  are characterized by a gapped bulk spectrum and the existence of protected gapless edge modes. Of particular interest are 1D topological superconductors~\cite{kitaev2001unpaired,Alicea2012,Beenakker2013}, where the manifestation of these modes are localized Majorana bound states (MBS). Due to the topological protection and non-Abelian exchange statistics of MBS~\cite{ivanov2001non,stern2004geometric,nayak2008non,alicea2011non}, topological superconductors might have applications in the field of quantum information processing~\cite{Kitaev2003}. Recently, it was proposed to realize this exotic phase in a semiconductor nanowire with strong spin-orbit coupling (SOC) under an applied Zeeman field, proximity-coupled to an {\it s}-wave superconductor~\cite{lutchyn2010majorana,oreg2010helical}. This was followed by several experimental studies showing evidence consistent with the existence of MBS, including a zero-bias peak in the differential conductance~\cite{mourik2012signatures,Das2012zero,deng2012anomalous,churchill2013superconductor}, and fractional Josephson effect~\cite{Rokhinson2012fractional}.

While the described system 
requires the breaking of time-reversal (TR) symmetry, it has been pointed out that MBS can also appear in a TR-invariant system, at the end of a time-reversal invariant topological superconductor~\cite{qi2009time} (TRITOPS). Intuitively, this phase can be thought of as two copies (related by TR) of a spinless {\it p}-wave superconductor. In that sense, the TRITOPS is to the spinless {\it p}-wave superconductor what the topological insulator is to the quantum Hall state.

 A one-dimensional TRITOPS has a pair of MBS at each end, as dictated by Kramers' theorem. Even though two MBS can usually be viewed as a normal fermion, it is predicted that such a Kramers pair will in fact have unusual properties, distinguishing it from a trivial fermion~\cite{liu2014non,Keselman2013inducing}.
 A number of models supporting time-reversal invariant topological superconductivity have been proposed recently~\cite{Keselman2013inducing,Fu2010,Deng2012,Nakosai2012topological,Seradjeh2012majorana,Wong2012,Zhang2013time,Nakosai2013,gaidamauskas2014majorana}. 

In this paper we suggest a mechanism for realizing a TRITOPS 
 due to repulsive interactions. Unlike previous proposals, this mechanism does not rely on the use of unconventional superconductors, and does not require the application of magnetic field or fine tuning of any parameter.

\begin{figure}
\includegraphics[clip=true,trim =0cm 6cm 0cm 4cm,width=0.45\textwidth]{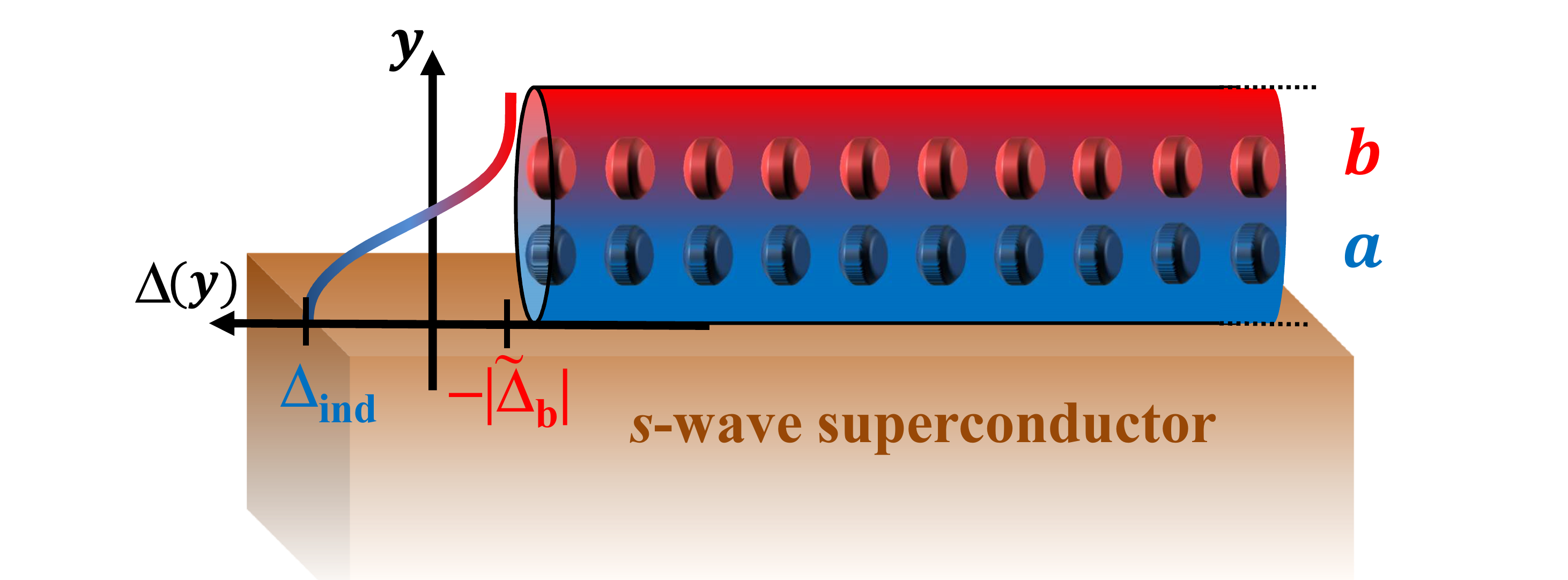}
\caption{The proposed system consists of a single quasi-1D wire (modeled by two chains) with SOC, coupled to a conventional {\it s}-wave superconductor. Integrating out the degrees of freedom of the superconductor generates a pairing potential $\Delta_{\rm{ind}}$ on the chain adjacent to the superconductor. Repulsive interactions in the wire which resist local pairing of electrons, induce a pairing potential $\tilde\Delta_b$ on the "{\it b}" chain with an opposite sign to $\Delta_{\rm ind}$.}\label{fig:setup}
\end{figure}

 The proposed system is shown schematically in Fig.~\ref{fig:setup}. A quasi 1D wire with SOC is proximity coupled to a conventional {\it s}-wave superconductor.
 In such a system, we show that 
  short range repulsive interactions in the wire counteract the proximity effect of the bulk superconductor, thereby inducing an internal $\pi$ junction~\cite{DeGennes1964Boundary} in the wire. The system can then be described in terms of two coupled single-mode wires 
   with pairing potentials of opposite signs. As was recently shown, 
such a setup can, under certain conditions, give rise to TRITOPS~\cite{Keselman2013inducing}. It should be stressed that since the $\pi$ junction naturally forms due to interactions, no fine-tuning of superconducting phases is required. 

Below, we use Hartree-Fock theory to show how the $\pi$ junction is created. We formulate the topological invariant that characterizes the TRITOPS phase and use it to map out the phase diagram.
We then move on to study the system using the density-matrix renormalization group (DMRG)~\cite{White1992PRL,White1992PRB}, and verify that the model indeed gives rise to a TRITOPS phase. Finally, we study the behavior of the system under breaking of TR symmetry, and demonstrate that the dependence of the splitting of the zero-bias conductance peak on the direction of the Zeeman field can be used as a distinct experimental signature of the TRITOPS phase.

\emph{The model.} We consider a quantum wire proximity coupled to an {\it s}-wave superconductor. The wire supports two modes, and has spin-orbit coupling and repulsive Coulomb interactions (assumed to be short ranged due to screening from the adjacent superconductor)
. The system is described by the following Hamiltonian:
\begin{equation}
\begin{array}{c}
H=\frac{1}{2}\displaystyle\sum\limits_k \Psi_k^\dag\mathcal{H}_{0,k}\Psi_k + \displaystyle\sum\limits_{i,\sigma} U_{\sigma}\hat{n}_{i\sigma\uparrow}\hat{n}_{i\sigma\downarrow}\\
\begin{split}
\mathcal{H}_{0,k}=&\left[\bar{\xi}_k+\delta\xi_k\sigma^z-\left(\bar{\alpha}+\delta\alpha\sigma^z\right)\sin{k}~s^z +t_{ab}\sigma^x\right]\tau^z\\
+&\Delta_{\rm ind}/2\left(1+\sigma^z\right)\tau^x ,
\end{split}
\end{array}\label{eq:Hamiltonian}
\end{equation}
where $\Psi_k^\dag=(\begin{matrix}\psi^\dag_k,&-i\psi^t_{-k}s^y\end{matrix})$. We model the wire using two chains labeled $a$ and $b$~\cite{chain_channel_note} 
, such that $\psi_k^\dag=(\begin{matrix}c_{a,k\uparrow}^\dag&c_{b,k\uparrow}^\dag&c_{a,k\downarrow}^\dag&c_{b,k\downarrow}^\dag\end{matrix})$. Diagonalizing $\mathcal{H}_{0,k}$ gives rise to two transverse subbands. $\vec{\tau}, \vec{\sigma}$, and $\vec{s}$ are Pauli matrices operating on particle-hole (PH), chain and spin degrees of freedom, respectively. Here, $\bar\xi_k, \delta\xi_k, \bar\alpha$ and $\delta\alpha$ are defined as  $(\xi_{k,a}\pm\xi_{k,b})/2$ and $(\alpha_a\pm\alpha_b)/2$, respectively, and $\xi_{k,\sigma}=2t_\sigma\left(1-\cos{k}\right)-\mu_\sigma$, $\sigma=a,b$. The parameters $t_\sigma, \alpha_\sigma, \mu_\sigma$ and $U_\sigma$ represent the hopping, SOC, chemical potential and on-site repulsion on each of the chains, while $t_{ab}$ is the hopping between the chains. The operator $\hat{n}_{i,\sigma,s}$ represents the number of particles with spin $s$ on site $i$ of chain $\sigma$. We note that $\mathcal{H}_{0,k}$ by itself cannot give rise to TRITOPS~\cite{Zhang2013time}, making the repulsive interaction term essential. 

\emph{Hartree-Fock analysis.} We consider a set of trial wave functions which are ground states of the following quadratic Hamiltonian:
\begin{equation}
\begin{split}
&H_{\rm \scriptscriptstyle HF}=\frac{1}{2}\displaystyle\sum\limits_k \Psi_k^\dag\mathcal{H}_k^{\rm \scriptscriptstyle HF}\Psi_k,\\
\mathcal{H}_k^{\rm \scriptscriptstyle HF}=&\tilde{\mathcal{H}}_{0,k}+\tilde{\Delta}_b/2\left(1-\sigma^z\right)\tau^x ,\label{eq:H_HF}
\end{split}
\end{equation}
where $\tilde{\mathcal{H}}_{0,k}$ has the same form as $\mathcal{H}_{0,k}$, with effective parameters $\tilde\mu_a, \tilde\mu_b$ and $\tilde\Delta_{\rm ind}$, while $\tilde\Delta_b$ is an effective pairing potential on chain {\it b}. Due to the repulsive interactions in the wire, $\tilde\Delta_b$ will turn out to have an opposite sign with respect to $\Delta_{\rm ind}$. We choose the four effective parameters such that they minimize the expectation value of the full Hamiltonian,
\begin{equation}
\begin{split}
&\langle H\rangle_{_{\rm HF}}=E_0+\frac{1}{L}\sum_\sigma U_\sigma\left(N_{\sigma,\uparrow}N_{\sigma,\downarrow}+\left|P_\sigma\right|^2\right)\\
&\begin{array}{c c}
N_{\sigma,s}=\sum_k \langle c_{\sigma,k,s}^\dag c_{\sigma,k,s}\rangle_{_{\rm HF}} , & P_\sigma=\sum_k \langle c_{\sigma,k,\uparrow}^\dag c_{\sigma,-k\downarrow}^\dag\rangle_{_{\rm HF}}
\end{array}\\
&E_0=\frac{1}{2}\sum_{k,m,n}\mathcal{H}_{0,k,mn}\langle \Psi_{k,m}^\dag\Psi_{k,n}\rangle_{_{\rm HF}}
\end{split}\label{eq:H_avg}
\end{equation}
where $L$ is the number of sites in each chain, and we have used Wick's theorem, noting the exchange term vanishes due to the $s^z$ conservation of $\mathcal{H}_k^{\rm \scriptscriptstyle HF}$. For details of the calculation see the Supplemental Material (SM) ~\cite{SM}.

We are interested in the conditions under which $\mathcal{H}_k^{\rm \scriptscriptstyle HF}$ is in the topological phase. This Hamiltonian possesses both TR symmetry $\Theta=is^yK$ and PH symmetry $\Xi=\tau^ys^yK$, expressed by $\Theta\mathcal{H}_k^{\rm \scriptscriptstyle HF}\Theta^{-1}=\mathcal{H}_{-k}^{\rm \scriptscriptstyle HF}$ and $\Xi\mathcal{H}_k^{\rm \scriptscriptstyle HF}\Xi^{-1}=-\mathcal{H}_{-k}^{\rm \scriptscriptstyle HF}$, making it in symmetry class DIII~\cite{Altland1997} with a $\mathbb{Z}_2$ topological invariant in 1D~\cite{schnyder2008classification,kitaev2009periodic}. To obtain an expression for the $\mathbb{Z}_2$ invariant~\cite{Qi2010,budich2013topological} for our system, we use the chiral symmetry $\left\{\mathcal{H}^{\rm \scriptscriptstyle HF},\tau^y\right\}=0$, to divide $\mathcal{H}^{\rm \scriptscriptstyle HF}$ into two off-diagonal blocks
\begin{equation}
e^{i(\pi/4)\tau^x}\mathcal{H}^{\rm \scriptscriptstyle HF}_ke^{-i(\pi/4)\tau^x}=\begin{pmatrix} 0 & \mathcal{B}_k \\ \mathcal{B}_k^\dag & 0 \end{pmatrix}.
\end{equation}
Due to the additional $s^z$ symmetry of the Hamiltonian, $\mathcal{B}_k$ can be further separated into two diagonal blocks, $\mathcal{B}^\uparrow_k$ and $\mathcal{B}^\downarrow_k=\mathcal{B}^\uparrow_{-k}$. The $\mathbb{Z}_2$ invariant is then given~\cite{SM} by the parity of the winding number of $\theta_k$, defined by $\exp\left(i\theta_k\right)=\det{\mathcal{B}^\uparrow_k}/|\det{\mathcal{B}^\uparrow_k}|$. For our model one has
\begin{equation}
\det{\mathcal{B}^\uparrow_k} = t_{ab}^2+\tilde\Delta_{\rm ind}\tilde\Delta_b -\tilde\varepsilon_{a,k}\tilde\varepsilon_{b,k}-i(\tilde\Delta_{\rm ind}\tilde\varepsilon_{b,k}+\tilde\Delta_b\tilde\varepsilon_{a,k}) ,\label{eq:Z2}
\end{equation}
where $\tilde\varepsilon_{\sigma,k}=2t_\sigma(1-\cos{k})-2\alpha_\sigma\sin{k}-\tilde\mu_\sigma$. It can be shown from Eq.~\eqref{eq:Z2} that in order to have a non trivial winding number (i.e. odd), one must have different SOC on the two chains, $\alpha_a/t_a\neq \alpha_b/t_b$. We note, however, 
that this requirement can be relaxed by adding a SOC term associated with hopping between the chains.  We can now solve the Hartree-Fock problem for the effective parameters, and then calculate the $\mathbb{Z}_2$ invariant using Eq.~\eqref{eq:Z2} to obtain the phase diagram of the system.

In addition to the trivial and the topological superconducting phases other competing phases may appear, which are not accounted for in our trial wave functions.
In particular, absent from Eq.~\eqref{eq:H_HF} are terms which break the lattice translational invariance and drive the system to a spin-density wave (SDW) state~\cite{Footnote}. To examine this possibility, after obtaining the effective parameters in Eq.~\eqref{eq:H_HF}, we add to $H_{\rm \scriptscriptstyle HF}$ the term $-\sum_{\sigma,q,s}\phi_{\sigma -q s}\hat{\rho}_{\sigma q s}$, where $\hat{\rho}_{\sigma q s}$ is the Fourier transform of $\hat{n}_{i,\sigma s}$. We then use {\it linear response} to
calculate the Hessian matrix $\partial^2{\langle H\rangle_{_{\rm HF}}}/\partial\phi_{\sigma',q,s'}\partial\phi_{\sigma,-q,s}$. For the Hartree-Fock solution to be locally stable to formation of SDW we demand the Hessian to be positive definite~\cite{SM}.

\begin{figure}
\includegraphics[clip=true,trim =0.5cm 0cm 0.6cm 0.95cm,width=0.35\textwidth]{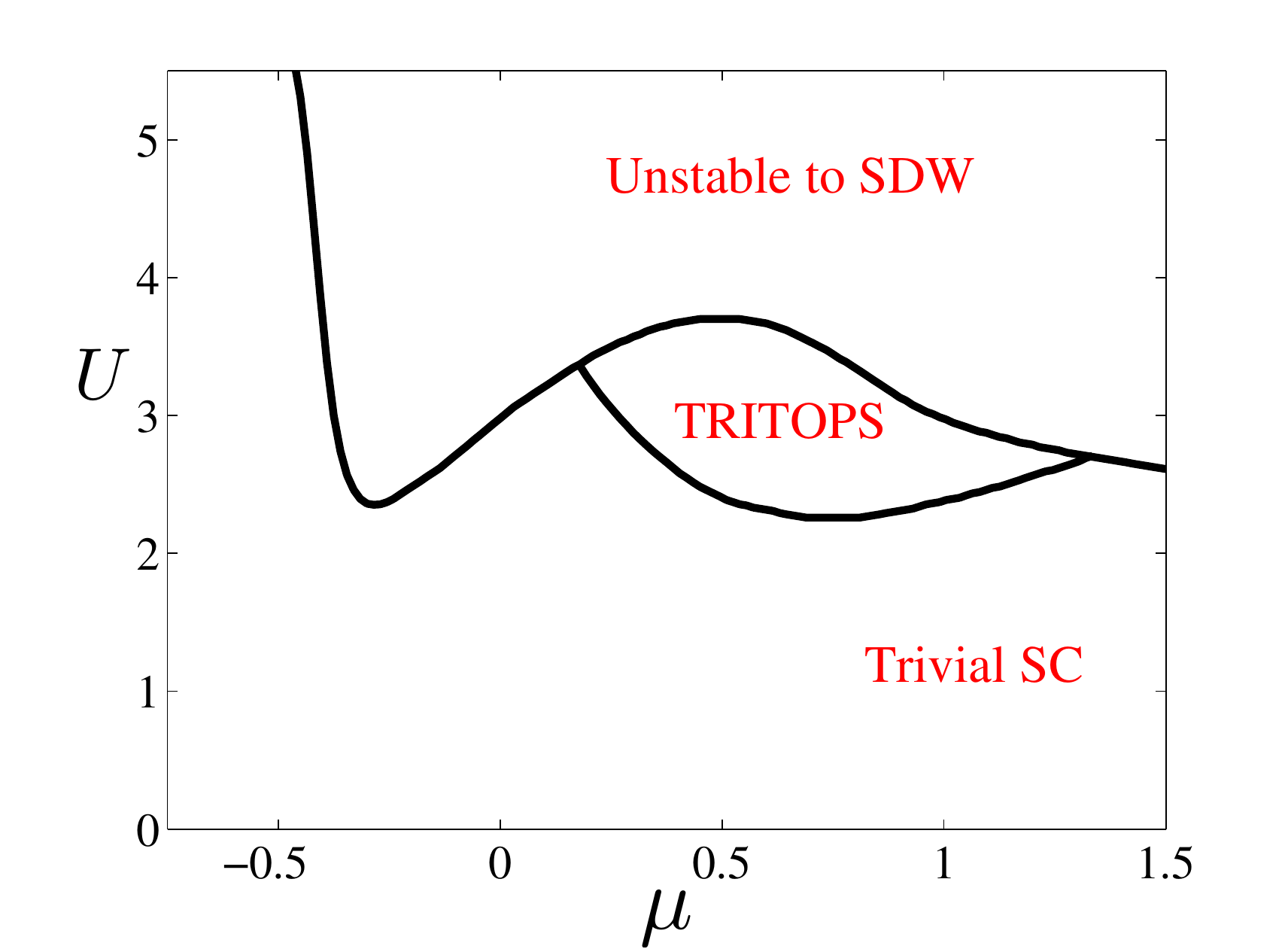}\label{fig:phase_diagram}
\caption{Hartree-Fock phase diagram as a function of chemical potential $\mu_a=\mu_b=\mu$, and interaction strength $U_a=U_b=U$, for $t_a=t_b=1, t_{ab}=0.4, \alpha_a=0, \alpha_b=0.6$
, $\Delta_{\rm ind}=1$ and $L=200$. The diagram includes a TRITOPS phase, a trivial superconductor phase, and a region in which the Hartree-Fock solution is locally unstable to the formation of spin-density waves~\cite{Footnote}.}\label{fig:phase_diagram}
\end{figure}
In Fig.~\ref{fig:phase_diagram} we present the Hartree-Fock phase diagram as a function of chemical potential and interaction strength for a specific set of wire parameters. In the SM~\cite{SM} we present results for different sets of parameters.

\emph{DMRG analysis.} We next verify the appearance of the topological phase using DMRG analysis of the model in Eq.~\eqref{eq:Hamiltonian}.
As was already mentioned, the TRITOPS hosts two Majorana
modes, related by time-reversal operation, at each end of the wire. We denote by $\gamma_{L(R)}$ one
Majorana operator localized on the left (right) end of the wire, and by $\tilde{\gamma}_{L(R)}$ its time-reversed partner.
These four Majorana operators give rise to two zero-energy fermionic operators $f_{L,R}=\gamma_{L,R}+i\tilde{\gamma}_{L,R}$.
Denote by $\left|\Psi\right\rangle$ the many body ground state
of the system in which both of these fermionic states are unoccupied $n_{f_R}=n_{f_L}=0$, where
$n_{f_{R(L)}}=\left\langle\Psi\right|f_{R(L)}^{\dagger}f_{R(L)}\left|\Psi\right\rangle$.
It is clear then that the four states $\left|\Psi\right\rangle $, $f_{L,R}^{\dagger}\left|\Psi\right\rangle $ and
$f_{L}^{\dagger}f_{R}^{\dagger}\left|\Psi\right\rangle $ are all degenerate in the thermodynamic limit.
The fourfold degeneracy of the ground state is a distinct signature
of the TRITOPS phase easily accessible using DMRG~\cite{Stoudenmire2011}.
Moreover, considering even and odd fermion parity sectors separately, we expect a double degeneracy in each.
In a finite system, a non-vanishing overlap between the Majorana modes living on opposite ends of the wire will give rise to an energy splitting exponentially decreasing with the system size.
However, the two odd fermion parity states will remain exactly degenerate for any system size due to Kramers' theorem.

\begin{figure}
    \subfloat[]{\includegraphics[clip=true, trim =-1cm 0cm 0cm 0cm,scale=0.35]{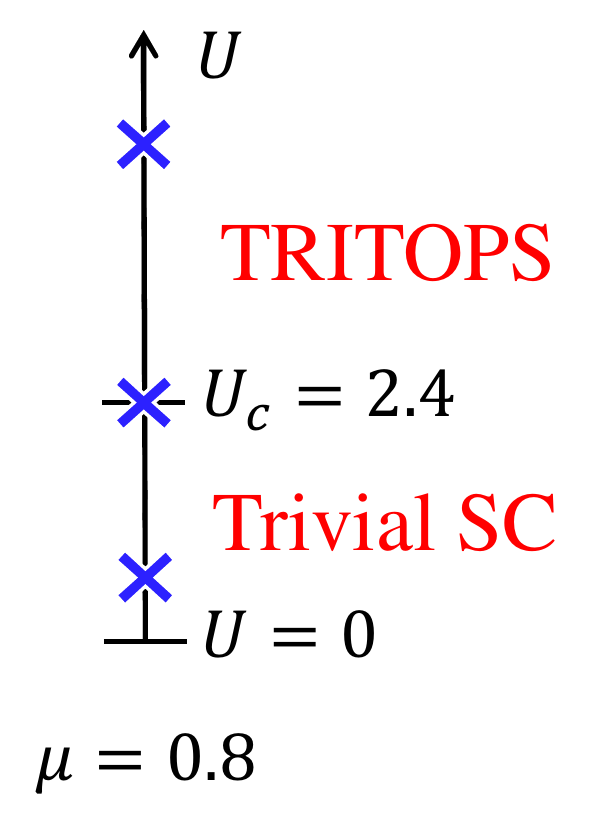}\label{fig:PhaseDiagram1D}}
    \subfloat[Trivial SC]{\includegraphics[clip=true, trim =-4cm 0cm 0.3cm 0.4cm,width=0.3\textwidth]{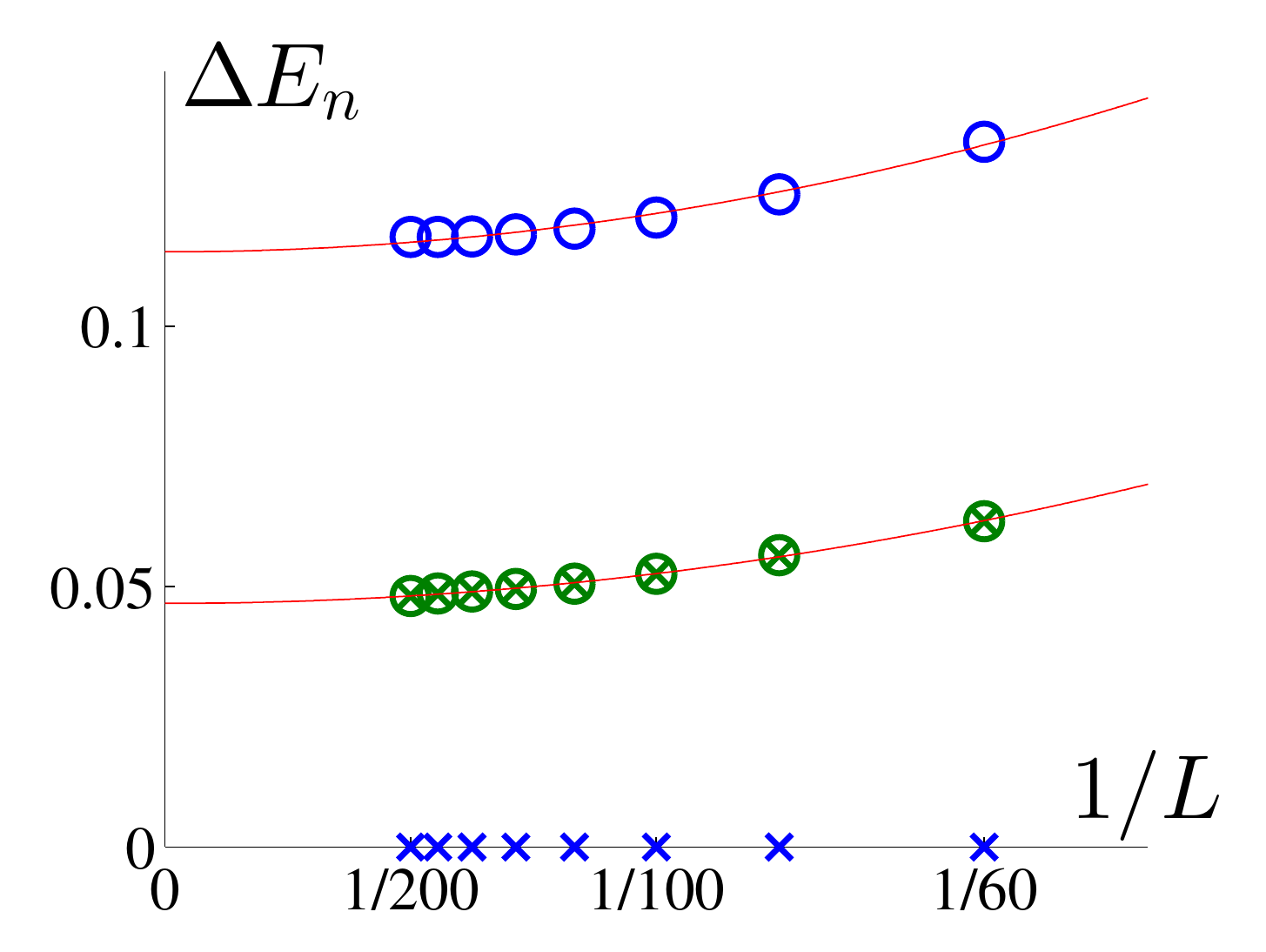}\label{fig:EnTrivial}}

    \subfloat[Critical Point]{\includegraphics[clip=true, trim =0cm 0.25cm 0cm 0cm,width=0.23\textwidth]{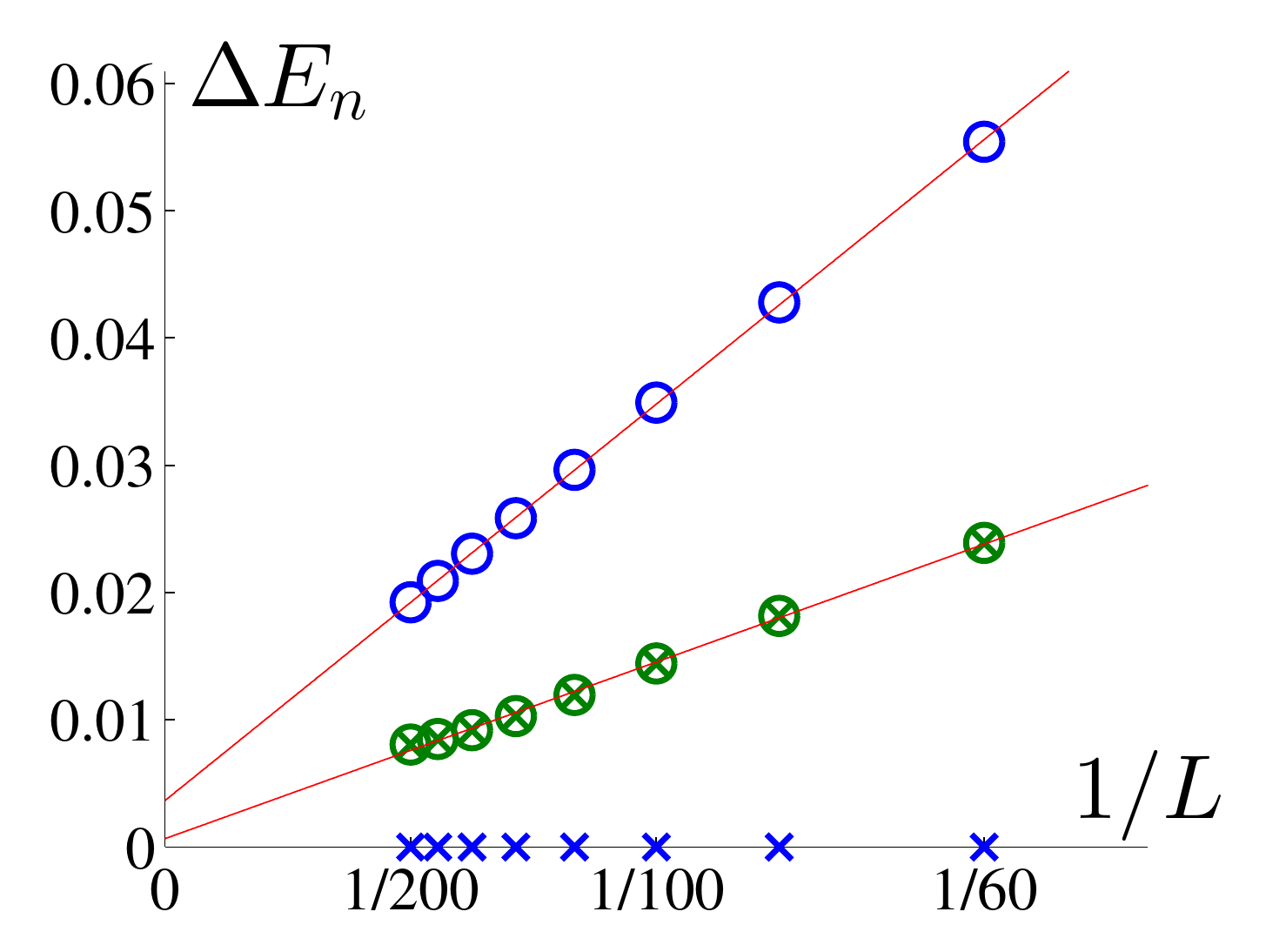}\label{fig:EnGapless}}
    \subfloat[TRITOPS]{\includegraphics[width=0.25\textwidth]{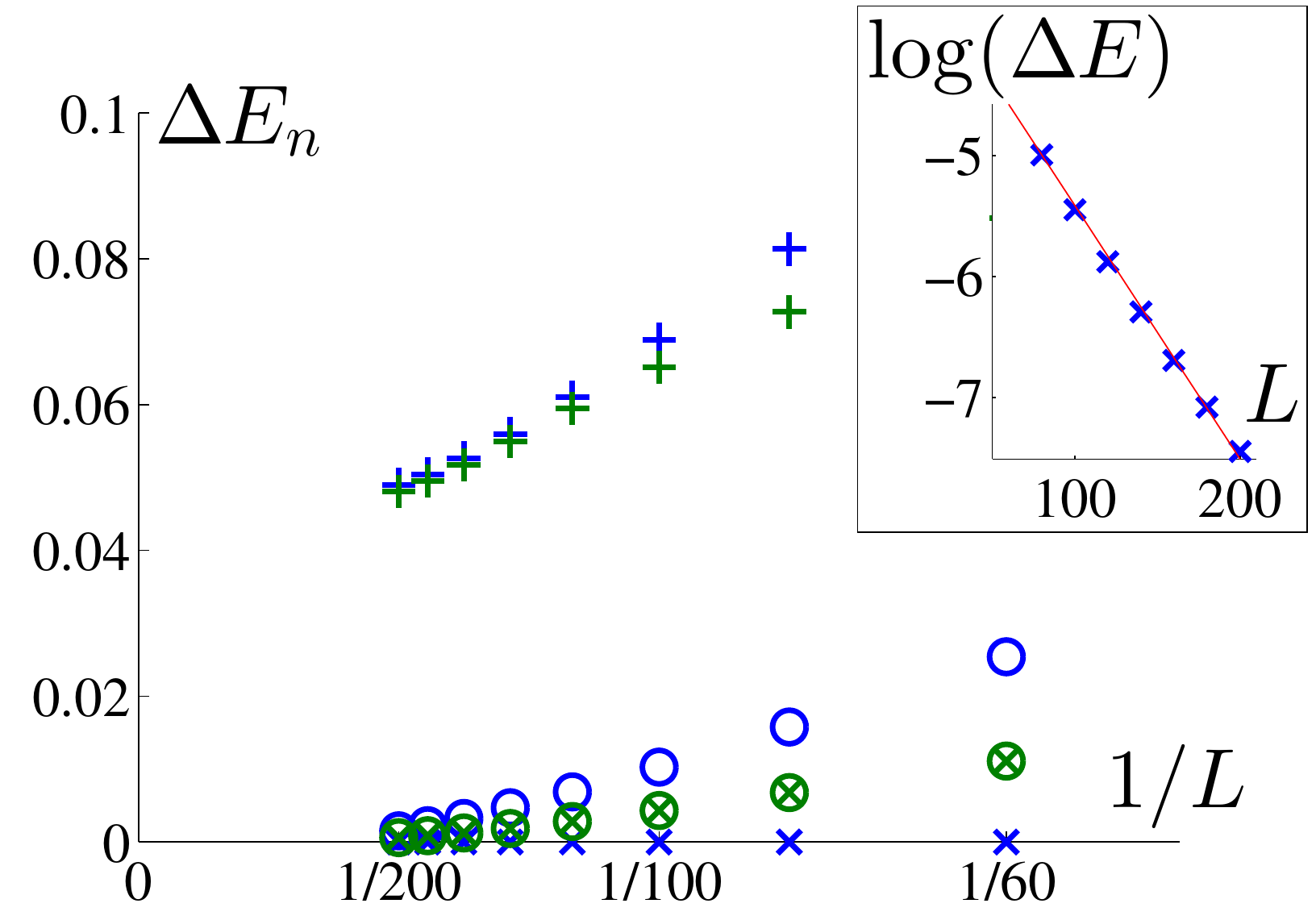}\label{fig:EnTopological}}

\caption{(a) Phase diagram obtained using DMRG. The system is in the trivial superconducting phase for $U<U_c$, and in the TRITOPS phase for $U>U_c$.
The system's parameters are $t_{a}=t_{b}=1,\ t_{ab}=0.4,\ \alpha_{a}=0,\ \alpha_{b}=0.6,\ \Delta_{\rm{ind}}=1,\ \mu=0.8$.
The low-lying energy spectrum of the system vs. $1/L$, where $L$ is the length of the wire, for the three marked points is plotted in (b)-(d).
Energies plotted in blue (green) correspond to energy states in the even (odd) fermion parity sectors. All energies are plotted with respect to the energy of the ground state, which is in all cases the lowest energy state in the even fermion parity sector, $\Delta E_n=E_n-E^{\rm{even}}_0$.
(b) $U=0.5<U_c$. The system is in the trivial superconducting phase. The first two states in the even and odd fermion parity sectors are shown. The ground state is unique and the gap tends to a constant as $L\rightarrow\infty$, with a quadratic correction in $1/L$ as expected (the red lines are quadratic fits). The first excited state which lies in the odd parity sector is doubly degenerate as expected from Kramers' theorem.
(c) $U=U_c=2.4$. Phase transition point. Once again, the first two states in each fermion parity sector are shown. All gaps scale linearly with $1/L$ in agreement with the system being gapless in the infinite size limit (the red lines are linear fits).
(d) $U=5.5>U_c$. The system is in the TRITOPS phase. Here, the three lowest states in each fermion parity sector are shown. 
The result is consistent with a fourfold degenerate ground state in the thermodynamic limit, separated by a finite gap from the rest of the spectrum.
The inset shows the energy difference $\Delta E$ between the lowest states in the even and odd fermion parity sectors on a semilogarithmic scale as a function of $L$. The result is consistent with an exponential dependence of $\Delta E$ on the system size.}
\end{figure}

A phase diagram obtained using DMRG is shown in Fig. \ref{fig:PhaseDiagram1D}.
Keeping the chemical potential $\mu$ constant we vary the on-site repulsive interaction strength $U$.
At $U=0$ the system is in a trivial superconducting phase with a finite gap for single particle excitations.
At a critical interaction strength $U_c$ a phase transition occurs and the gap closes.
For $U>U_c$ the gap reopens with the system now being in the TRITOPS phase.

To obtain the phase diagram we calculate the lowest energy states and analyze the scaling of the gaps in the system as we increase its size.
We take advantage of fermion parity conservation and calculate the energies in the even and odd fermion parity sectors separately.
A detailed explanation of the analysis is given in the SM~\cite{SM}.
In Figs. \ref{fig:EnTrivial}-\ref{fig:EnTopological} we present the scaling of the low-lying energy spectrum with the length of the wire
at three different points in the phase diagram: one in the trivial superconducting phase, one in the TRITOPS phase and one at the critical point where the gap closes.
In the trivial superconducting phase (Fig. \ref{fig:EnTrivial}), we observe a unique ground state as expected. The gap to the first excited
state extrapolates to a finite value in the limit of an infinite system.
Note that this state is doubly degenerate due to Kramers' theorem, as it is in the odd fermion parity sector. The gap to the first excited state in the even fermion parity sector is nearly twice as large, as expected.
At the phase transition, the gap closes. For a finite 1D system this means that the gaps should be inversely proportional to the size of the system, as can be clearly seen in Fig. \ref{fig:EnGapless}.
In the TRITOPS phase (Fig. \ref{fig:EnTopological}) the ground state is fourfold degenerate up to finite-size splittings.
The exponential dependence of the energy splitting on the length of the wire can be clearly seen from the inset.
The two lowest energy states in the odd fermion parity sector indeed remain degenerate for any system size.
Excited levels are separated from the ground state manifold by a finite gap.

We thus conclude that the DMRG study supports the Hartree-Fock analysis of the system confirming the appearance of the TRITOPS phase due to repulsive interactions. Interestingly, the range of stability of the TRITOPS phase obtained by DMRG persists to values of $U$ higher than those suggested by the Hartree-Fock calculation for the same parameters (cf. Fig.~\ref{fig:phase_diagram} and the SM~\cite{SM}).  

\emph{Experimental signature and breaking of TR invariance.} In the TRITOPS phase the wire supports two MBS at each end. Measurement of differential conductance, through a lead coupled to the end of the wire, should therefore reveal those states via a peak at zero bias. At $T\rightarrow 0$ the height of the peak should be quantized to $4e^2/h$, as opposed to the TR-broken topological superconductors in which the peak is generally quantized to $2e^2/h$~\cite{Bolech2007Observing,Law2009majorana,akhmerov2011quantized}.

\begin{figure}
\subfloat[]{\includegraphics[clip=true,trim =3.1cm 8.75cm 4cm 9.4cm,width=0.215\textwidth]{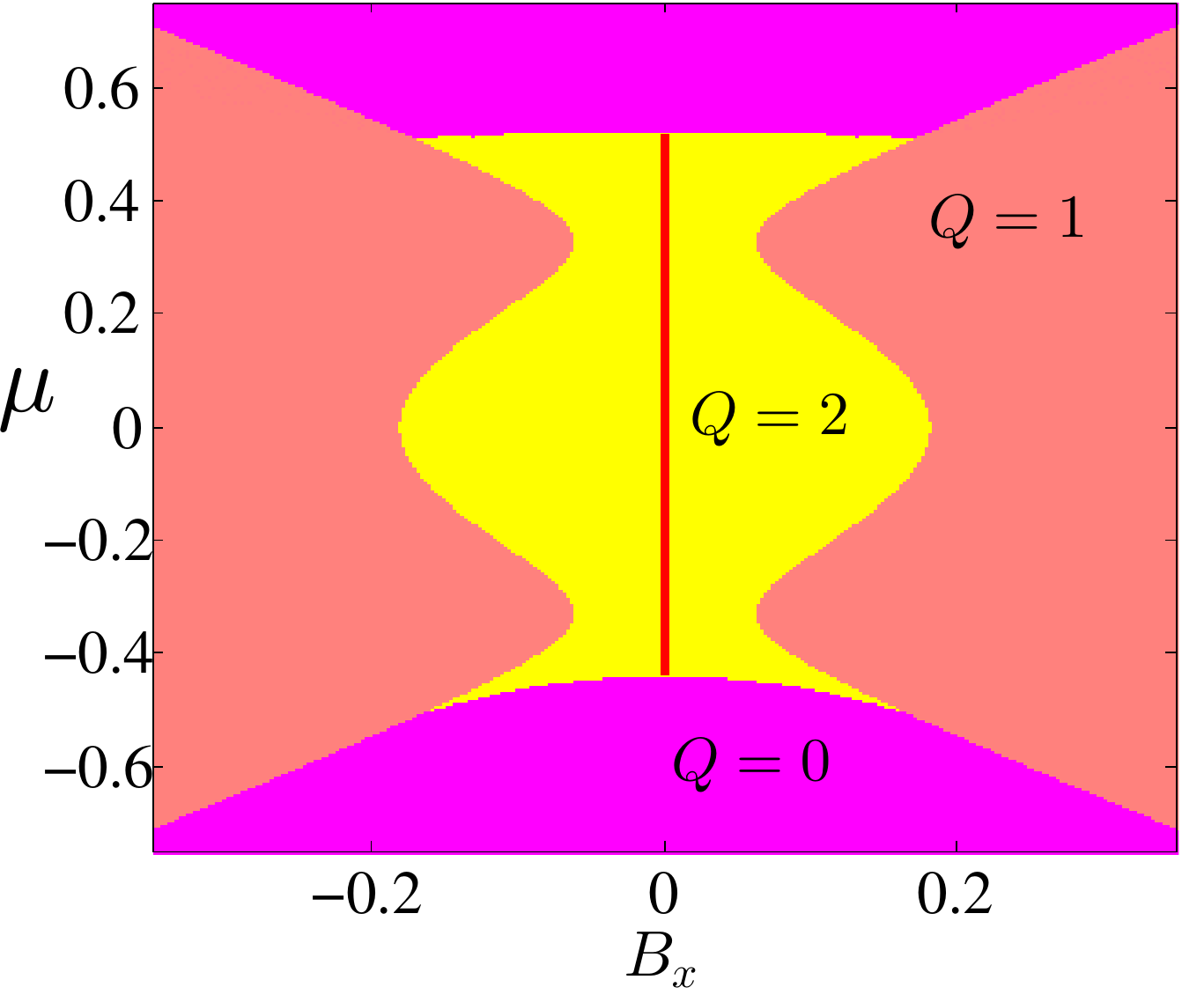}\label{fig:BDI_Z_invariant}}
\subfloat[]{\includegraphics[clip=true,trim =0cm -0.2cm 0cm 0.8cm,width=0.256\textwidth]{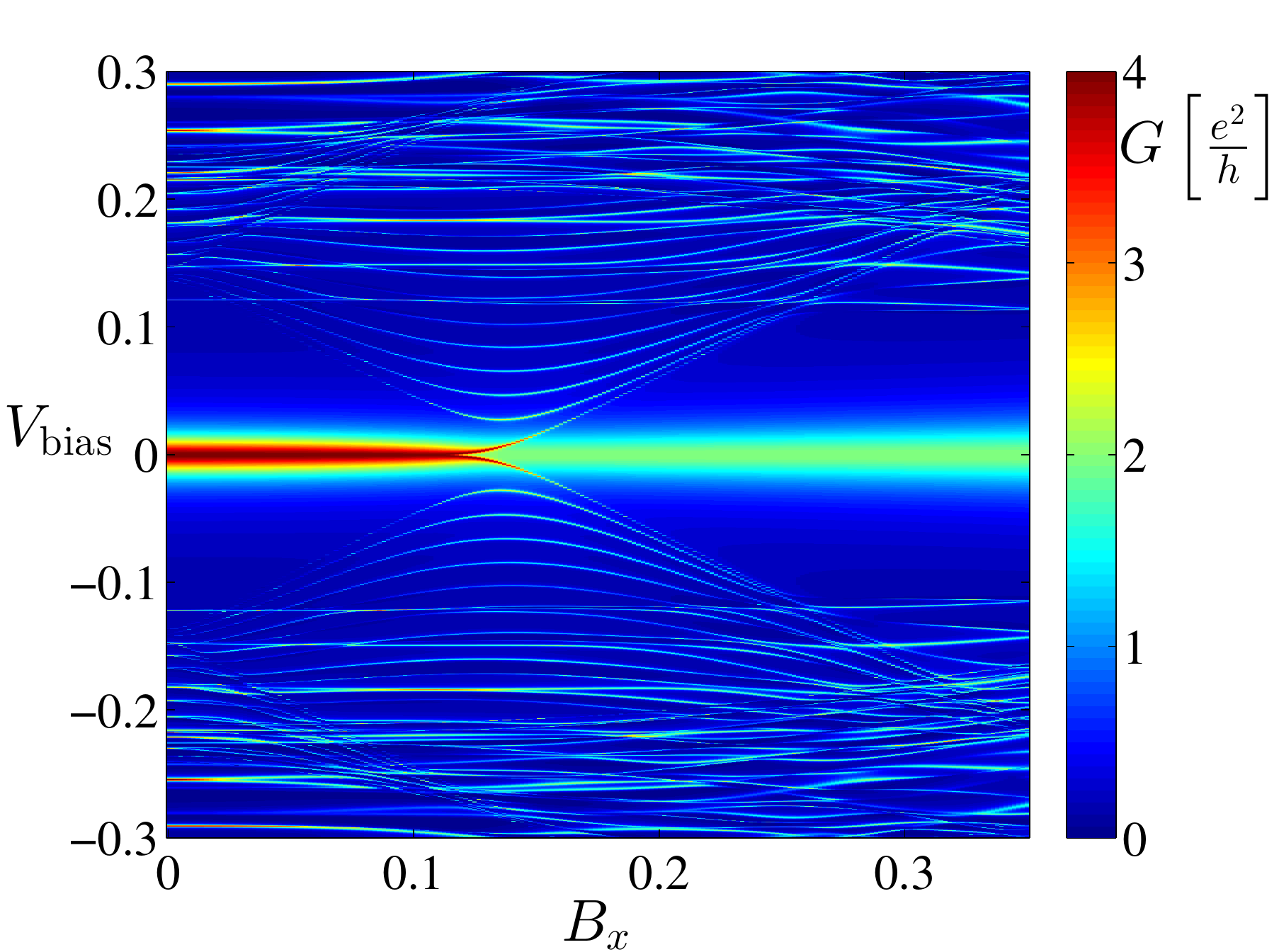}\label{fig:differential_conductance}}
\caption{(a) 
Phase diagram of the Hartree-Fock Hamiltonian of Eq.~\eqref{eq:H_HF} as function of chemical potential $\mu_a=\mu_b=\mu$, and Zeeman field along $x$ (perpendicular to the SOC), for $t_a=t_b=1,t_{ab}=0.4, \alpha_a=0, \alpha_b=0.6, \tilde\Delta_{\rm ind}=0.3$ and $\tilde\Delta_b=-0.15$. The system is in symmetry class BDI, and is characterized by a $\mathbb{Z}$ invariant, $Q$. 
$Q$ equals the number of MBS at each end of the wire. The TRITOPS phase is marked by a red line. (b) Differential conductance through a single lead connected to the wire as a function of bias voltage and Zeeman field for $\mu=0.15$. A wire of $L=100$ sites was used in the calculation.}
\end{figure}

To explore this difference we can study the behavior of the system under the breaking of TR symmetry, by introducing a uniform magnetic field via the Zeeman term $\mathcal{H}_B=-\vec{B}\cdot \vec{s}$. When the magnetic field is applied parallel to the direction of the SOC ($z$ in our setup), the zero-bias peak (ZBP) splits linearly with the magnetic field, as the MBS are no longer protected by TR symmetry. In contrast, we can apply the field perpendicular to the SOC, e.g. along $x$. 
Even though TR symmetry is now broken, the Hamiltonian still has an antiunitary symmetry $\Lambda=s^xK$, expressed by $\Lambda\mathcal{H}_{k}\Lambda^{-1}=\mathcal{H}_{-k}$, which protects the MBS from splitting~\cite{tewari2012topological,Zhang2013time}. 
More specifically, due to this symmetry (together with PH symmetry) the Hamiltonian is in the BDI symmetry class~\cite{Altland1997} with a $\mathbb{Z}$ invariant, whose value determines the number of MBS at each end~\cite{schnyder2008classification,kitaev2009periodic}. In Fig.~\ref{fig:BDI_Z_invariant}, we plot the number of MBS as a function of chemical potential and Zeeman field as inferred from the BDI $\mathbb{Z}$ invariant, calculated according to Ref.~\cite{tewari2012topological}. One should note that in reality this symmetry is fragile, as it can be broken for instance by introducing a term $\alpha_{ab}\sigma^ys^x\tau^z$, which describes Rashba-type SOC associated with motion transverse to the wire. In such a case, however, we still expect the splitting at small magnetic fields to be of the form~\cite{Keselman2013inducing} $\vec{B}\cdot\hat{n}+O(B^3)$, for a certain unit vector $\hat{n}$, which depends on details of the SOC. 
This means that 
as long as $\vec{B}$ is applied perpendicular to $\hat{n}$, the Zeeman splitting of the ZBP scales as $B^3$ rather than linearly. This can serve as a distinct signature of the TRITOPS phase.

We next use the scattering matrix formalism to calculate the differential conductance through a single lead coupled to the system~\cite{blonder1982transition,shelankov1980resistance,SM}, described by the Hamiltonian of Eq.~\eqref{eq:H_HF} with an additional Zeeman field along $x$. As evident in Fig.~\ref{fig:differential_conductance}, there indeed exists a ZBP quantized to $4e^2/h$ which does not split at low magnetic fields. As the field is further increased, a topological phase transition occurs to a phase with a single MBS at each end, at which point the ZPB peak splits to three peaks. One of them stays at zero bias and is quantized to $2e^2/h$, while the other two become part of the bulk spectrum~\cite{Stanescu2012}.

\emph{Discussion.} We have demonstrated how the interplay between spin-orbit coupling and local repulsive interactions can lead to the formation of a TRITOPS phase in proximity-coupled quantum wires. The key for this mechanism is the fact that the induced gap function in a normal metal has the \emph{opposite sign} compared to that of the external superconductor, providing a $\pi$ phase shift necessary for the stabilization of the TRITOPS.

To estimate the strength of the repulsive interaction $U$ in a realistic setup, we take the wire diameter $d$, the spin-orbit length and the Fermi wavelength to be of order $100$ nm. This gives $U\sim e^2/\varepsilon d\sim1$--$10K$, for a reasonable dielectric constant of $\varepsilon\sim10$--$100$. Hence, $U$ can become comparable with other energy scales in the system such as the Fermi energy and $\Delta_{\rm ind}$, in agreement with our results for the condition to be in the TRITOPS phase. We note that one can realize this system also in a chain of quantum dots, thereby obtaining additional control over the system parameters~\cite{sau2012realizing,Fulga2013adaptive}, including $U$.

The most striking consequence of the TRITOPS phase is the appearance of a zero-bias peak in the tunneling conductance between a normal lead and the ends of the quantum wire at zero magnetic field. This peak is due to a zero-energy Kramers pair of Majorana end modes, protected by TR invariance. The peak has a unique behavior under the application of a magnetic field; the dependence of the splitting of the peak on the field direction can distinguish it from ordinary Andreev bound states.

\emph{Acknowledgments.} We would like to acknowledge discussions with L. Fu, A. Stern, K. Flensberg, F. von Oppen, F. Wang, J. Alicea, C. Marcus, Y. Schattner, and I. Seroussi. E.B. was supported by the Israel Science Foundation, by a Marie Curie CIG grant, and by the Robert Rees Fund. Y.O. was supported by an Israel Science Foundation grant, by the ERC advanced grant and by a DFG grant.

\bibliography{T_TRISC_References}

\begin{thebibliography}{50}
\bibitem [{Foo()}]{Footnote}%
  \BibitemOpen
  \href@noop {} {}\bibinfo {note} {This ordered phase is not permissible in the
  continuum limit in 1D, as it requires the breaking of a continuous symmetry.
  We therefore expect that long-range fluctuations (which are not accounted for
  in the HF treatment) would result in a Luttinger liquid phase.}\BibitemShut
  {Stop}%
\end{thebibliography}
\bibliographystyle{apsrev4-1}
\newpage

\begin{widetext}
\section{Supplementary Material}
\setcounter{equation}{0}
\subsection{Alternative Approach: Continuous Wire Model}

In the main text, we used a tight-binding model to describe the quantum wire. Below, we demonstrate that the mechanism for TRITOPS formation applies also
in a more general model for a wire with two transverse channels.

Consider a strip of width $d$ and length $L$. Before considering spin-orbit coupling, 
the spatial part of the wavefunctions of the two lowest energy transverse
modes is given by
\begin{equation}
\begin{split}
\Psi_{0,k}\left(x,y\right)=\sqrt{\frac{2}{d L}}\cos\left(\frac{\pi}{d}y\right)e^{ikx} \\
\Psi_{1,k}\left(x,y\right)=\sqrt{\frac{2}{d L}}\sin\left(\frac{2\pi}{d}y\right)e^{ikx},
\end{split}
\end{equation}
where $x$ denotes the direction along the wire and $y$ is the direction
perpendicular to it (see Fig. \ref{fig:TMSetup}). The origin is chosen such that the middle of the wire is at $y=0$. Note that each mode is doubly degenerate
due to the spin degree of freedom.

Next, we consider the effect of Rashba spin-orbit coupling using perturbation
theory. The Hamiltonian of the perturbation is
\begin{equation}
H_{SO}=-\lambda_{SO}kE_{y}\left(y\right)s_{z},
\end{equation}
where $s_{z}$ is a Pauli matrix acting in the spin subspace, $E_{y}\left(y\right)$
is the electric field in the $\hat{y}$ direction, and $\lambda_{SO}$
is the spin-orbit coupling strength.
Due to the confining potential of the wire, we expect $E_{y}\left(y\right)$
to be opposite in sign on the two sides of the wire. I.e., the function $E_{y}(y)$ has
an anti-symmetric component.

We now calculate the correction to the wavefunctions of the lowest energy modes due to the perturbation. The matrix element to the first excited state is diagonal in the spin
subspace, and is non-zero due to the anti-symmetric component of $E_{y}\left(y\right)$
\begin{equation}
\begin{split}
\Omega_{s,s'}=\left\langle \Psi_{1,k,s}\left|H_{SO}\right|\Psi_{0,k,s'}\right\rangle=
-s\delta_{s,s'}\frac{2 \lambda_{SO} k}{d}\int_{-d/2}^{d/2}dyE_{y}\left(y\right)\cos\left(\frac{\pi}{d}y\right)\sin\left(\frac{2\pi}{d}y\right)=\\
-s\delta_{s,s'}\frac{2 \lambda_{SO} k}{d}\int dyE_{y}^{\rm asym}\left(y\right)\cos\left(\frac{\pi}{d}y\right)\sin\left(\frac{2\pi}{d}y\right),
\end{split}
\end{equation}
where $s=\pm1$ for spin $\uparrow,\downarrow$ respectively, and $E_{y}^{\rm asym}=[E_y(y)-E_y(-y)]/2$.

For simplicity we take $E_{y}\left(y\right)=E_{y}^{\rm asym}\left(y\right)=E_{0}\sin\left(\frac{\pi}{d}y\right)$, and obtain
\begin{equation}
\Omega=-s\frac{2 \lambda_{SO} k}{d}E_{0}\int_{-d/2}^{d/2}dy \sin\left(\frac{\pi}{d}y\right) \cos\left(\frac{\pi}{d}y\right) \sin\left(\frac{2\pi}{d}y\right)=-s\frac{\lambda_{SO}}{2}kE_{0}.
\end{equation}

The perturbed wavefunctions of the lowest spin-degenerate modes are then
\begin{equation}
\Psi_{0,k,s}^{'}\left(x,y\right)\approx\Psi_{0,k,s}\left(x,y\right)+\frac{\Omega}{\Delta E}\Psi_{1,k,s}\left(x,y\right)=\sqrt{\frac{2}{d L}}\left[\cos\left(\frac{\pi}{d}y\right)-sk\frac{\lambda_{SO}E_{0}}{2\Delta E}\sin\left(\frac{2\pi}{d}y\right)\right]e^{ikx}\label{eq:PerturbedWavefunction}
\end{equation}
where $\Delta E$ is the energy difference between the two transverse modes in the wire.

As can be seen from Eq. \ref{eq:PerturbedWavefunction} the wavefunctions
of a right moving mode ($k=k_{F}>0$) with spin $\uparrow$ , as well as
of a left moving mode ($k=-k_{F}<0$) with spin $\downarrow$, are shifted in real space
to $y<0$. The wavefunctions corresponding to the second pair of Fermi points, a
right mover with spin $\downarrow$ and a left mover with spin $\uparrow$,
are shifted to $y>0$. The corresponding probability
distributions $\left|\Psi\left(y\right)\right|^{2}$ are plotted in
fig. \ref{fig:TMWavefunction}.

Assume the bulk superconductor is placed at $y<0$ as shown in Fig. \ref{fig:TMSetup},
and denote by $\Delta_{1,2}$ the induced superconducting pairing
felt by the modes corresponding to the two pairs of Fermi points. Due to the different
spatial extent of the modes (the former pair being on average closer to the superconductor than the latter) we expect the
strength of the induced pairing to be different with $\Delta_{1}>\Delta_{2}$.

\begin{figure}
\begin{centering}
\subfloat[\label{fig:TMSetup}]{\includegraphics[clip=true,trim =0cm 2cm 0cm 2cm,scale=0.35]{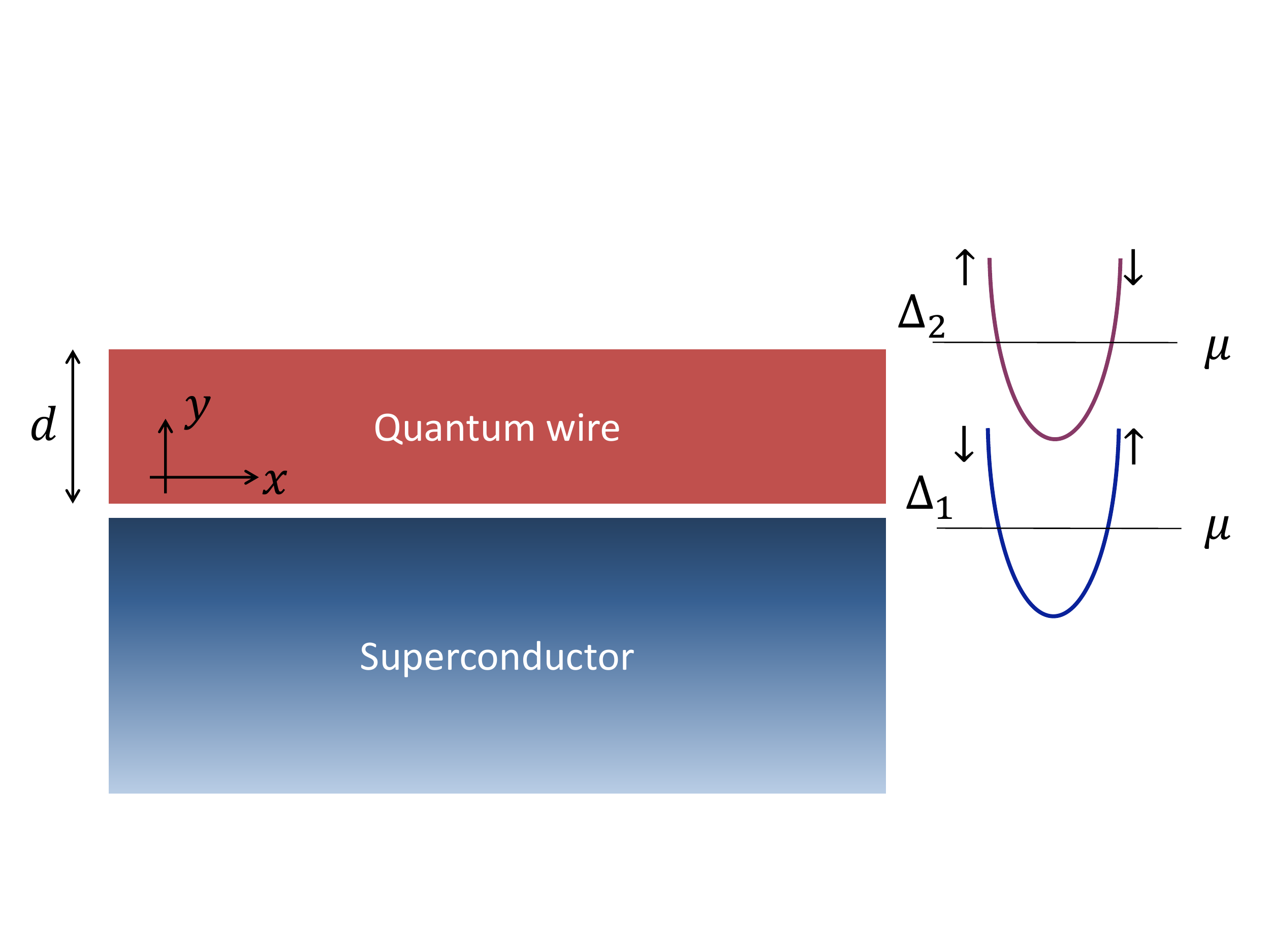}}
\subfloat[\label{fig:TMWavefunction}]{\includegraphics[scale=0.7]{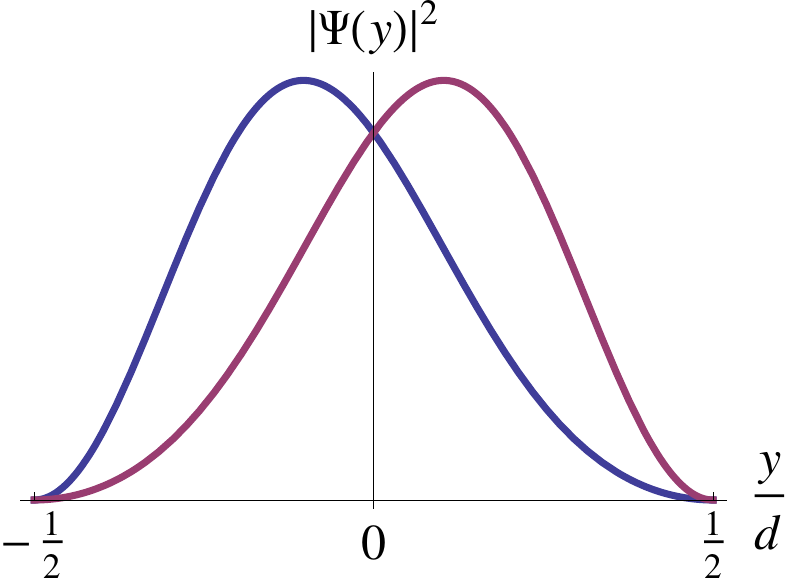}}
\par\end{centering}

\caption{(a) The setup considered consists of a wire of width $d$ on top of
a superconductor. Due to Rashba spin-orbit coupling, the eigenfunctions
for one pair of Fermi points are shifted to $y<0$ and have an induced
superconducting pairing $\Delta_{1}$, while the eigenfunctions for the other pair of Fermi
points are shifted to $y>0$ and have an induced pairing $\Delta_{2}<\Delta_{1}$.
(b) Probability distribution for the perturbed wavefunctions. The
wavefunctions of a rightmover with spin $\uparrow$ and of a leftmover
with spin $\downarrow$ are shifted to $y<0$ (blue curve), whereas
the wavefunctions of a rightmover with spin $\downarrow$ and a leftmover
with spin $\uparrow$ are shifted to $y>0$ (red curve).}
\end{figure}

Short-range repulsive interactions in the wire will then suppress the superconducting
pairing uniformly in $k$. Denote the suppressed pairing potentials
by $\tilde{\Delta}_{1,2}$. Following Qi et. al.~\cite{Qi2010}, a Fermi surface
$\mathbb{Z}_{2}$ invariant for a time-reversal invariant topological
superconductor in 1D can be written simply as the product of the signs
of the superconducting pairing, $\underset{s}{\prod}\rm{sign}\left(\Delta_{s}\right)$,
where $s$ is summed over all the Fermi points between 0 and $\pi$.
For strong enough interactions one can expect to have $\tilde{\Delta}_{1}>0>\tilde{\Delta}_{2}$
corresponding to a non-trivial $\mathbb{Z}_{2}$ index.

We therefore argue that repulsive interactions in a quantum wire with
spin-orbit coupling, proximity coupled to a conventional superconductor can naturally
give rise to a time-reversal invariant topological superconducting
phase.

\subsection{Hartree-Fock}

In real space basis the full Hamiltonian of the system is given by
\begin{equation}
\begin{split}
H=\displaystyle\sum_{n=1}^L& -\mu_\sigma c^\dag_{\sigma,n,s}c^{}_{\sigma,n,s} + w_\sigma e^{i\frac{\lambda_\sigma}{2}s^z_{ss'}}c^\dag_{\sigma,n+1,s}c^{}_{\sigma,n,s'} + t_{ab}\sigma^x_{\sigma,\sigma'}c^\dag_{\sigma,n,s}c^{}_{\sigma',n,s}+\\ +&\Delta_{\rm ind}c^\dag_{\sigma,n,\uparrow}c^\dag_{\sigma,n,\downarrow} +h.c. +U_\sigma c^\dag_{\sigma,n,\uparrow}c^{}_{\sigma,n,\uparrow}c^\dag_{\sigma,n,\downarrow}c^{}_{\sigma,n,\downarrow},
\end{split}
\end{equation}
where the parameters $w_\sigma$ and $\lambda$ are related to the hopping along the chains $t_\sigma$ and the SOC $\alpha_\sigma$ introduced in the main text by $t_\sigma=w_\sigma\cos(\lambda/2)$ and $\alpha_\sigma=w_\sigma\sin(\lambda/2)$.

Since we'll be treating a system with periodic boundary conditions we shall work with a Hamiltonian in momentum space representation, given in Eq. (1) of the main text and repeated below for clarity
\begin{equation}
\begin{array}{c}
H=\frac{1}{2}\displaystyle\sum\limits_k \Psi_k^\dag\mathcal{H}_{0,k}\Psi_k + \displaystyle\sum\limits_{i,\sigma} U_{\sigma}\hat{n}_{i\sigma\uparrow}\hat{n}_{i\sigma\downarrow}\\
\mathcal{H}_{0,k}=\left[\bar{\xi}_k+\delta\xi_k\sigma^z-\left(\bar{\alpha}+\delta\alpha\sigma^z\right)\sin{k}~s^z +t_{ab}\sigma^x\right]\tau^z+\Delta_{\rm ind}/2~\cdot\left(1+\sigma^z\right)\tau^x,
\end{array}\label{Hamiltonian}
\end{equation}
where $\Psi_k^\dag=(\begin{matrix}\psi^\dag_k,&-is^y\psi_{-k}\end{matrix})$.
The $s^z$ symmetry of the Hamiltonian allows us to write the non-interacting part using a $4\times4$ matrix rather than $8\times8$, in the following form:
\begin{equation}
\begin{split}
&H=H_0+H_{\rm I}\\
&\begin{array}{c c c} H_0=E_{\rm c}+\displaystyle\sum_k \Phi^\dag_k\bar{\mathcal{H}}_{0,k}\Phi_k&,&H_{\rm I}=\frac{1}{L}\displaystyle\sum_{k,p,q,\sigma} U_\sigma c^\dag_{\sigma,k+q\uparrow}c^\dag_{\sigma,p-q\downarrow}c_{\sigma,p\downarrow} c_{\sigma,k\uparrow}\end{array}
\end{split},
\end{equation}
where we define $\Phi^\dag_k=\begin{pmatrix} c^\dag_{a,k\uparrow} & c^\dag_{b,k\uparrow} & c_{a,-k\downarrow} & c_{b,-k\downarrow} \end{pmatrix}$
and
\begin{equation}
\begin{array}{ccc}
\bar{\mathcal{H}}_{0,k} =\begin{pmatrix}\varepsilon_{a,k}&t_{ab}&\Delta_{\rm ind}&0\\t_{ab}&\varepsilon_{b,k}&0&0\\ \Delta_{\rm ind}&0&-\varepsilon_{a,k}&-t_{ab}\\ 0&0&-t_{ab}&-\varepsilon_{b,k}\end{pmatrix}&,&E_{\rm c}=\displaystyle\sum_{\sigma,k}\varepsilon_{\sigma,k}.
\end{array}
\end{equation}
We distinguish between the $4\times4$ Hamiltonian $\bar{\mathcal{H}}_{0,k}$ and the $8\times8$ Hamiltonian ${\mathcal{H}}_{0,k}$ in Eq.~(\ref{Hamiltonian}).

We now want to consider a set of trial wave-functions for the ground state of the system. We choose these wave-functions to be the ground-states of the following quadratic Hartree-Fock Hamiltonian:
\begin{equation}
\begin{matrix}H^{\scriptscriptstyle{\rm HF}}=\displaystyle\sum_k \Phi^\dag_k\bar{\mathcal{H}}^{\scriptscriptstyle{\rm HF}}_k\Phi_k&,& \bar{\mathcal{H}}^{\scriptscriptstyle{\rm HF}}_k =\begin{pmatrix}\tilde\varepsilon_{a,k}&t_{ab}&\tilde\Delta_{\rm ind}&0\\t_{ab}&\tilde\varepsilon_{b,k}&0&\tilde\Delta_b\\ \tilde\Delta_{\rm ind}&0&-\tilde\varepsilon_{a,k}&-t_{ab}\\ 0&\tilde\Delta_b&-t_{ab}&-\tilde\varepsilon_{b,k}\end{pmatrix}\end{matrix}, \label{eq:HF_Hamiltonian}
\end{equation}
with effective parameters $\tilde\mu_a, \tilde\mu_b,\tilde\Delta_{\rm ind}$ and $\tilde\Delta_b$. We then calculate the expectation value of the full Hamiltonian in these ground states and determine the effective parameters as those for which this expectation value is minimized.

We begin by explicitly calculating the ground state of $H^{\scriptscriptstyle{\rm HF}}$. For this we need to diagonalize $\bar{\mathcal{H}}^{\scriptscriptstyle{\rm HF}}_k$ for each $k$. The Hamiltonian $\bar{\mathcal{H}}^{\scriptscriptstyle{\rm HF}}_k$ has a chiral symmetry given by $\tau^y$ and satisfying $\left\{\bar{\mathcal{H}}_k^{\scriptscriptstyle{\rm HF}},\tau^y\right\}=0$. It can therefore be brought to the following form by diagonalization
\begin{equation}
\begin{matrix}\bar{\mathcal{H}}^{\scriptscriptstyle{\rm HF}}_k=\mathcal{U}_k\mathcal{D}_k\mathcal{U}^\dag_k &,& \mathcal{D}_k=\begin{pmatrix}\epsilon_{1,k}& & &\\ & \epsilon_{2,k}& & \\ & & -\epsilon_{1,k}&\\ & & &-\epsilon_{2,k}\end{pmatrix}&,&\mathcal{U}_k=\begin{pmatrix}u(k)&v(k)\\ -v(k)&u(k)\end{pmatrix}\end{matrix},\label{eq:diagonalizing}
\end{equation}
where $\epsilon_{1,k},\epsilon_{2,k}$ are defined positive, and $u(k)$ and $v(k)$ are both $2\times2$ matrices operating on the chains degree of freedom ($\sigma=a,b$). The many body ground-state $\left|\Psi_{\scriptscriptstyle{\rm HF}}\right>$ of the Hartree-Fock Hamiltonian is the vacuum state of the fermionic operators given by
\begin{equation}
\begin{matrix}\Gamma^\dag_k=\Phi^\dag_k\mathcal{U}_k&,& \Gamma^\dag_k=\begin{pmatrix}\gamma^\dag_{{\rm I},k}&\gamma^\dag_{{\rm II},k}&\tilde\gamma^{}_{{\rm I},k}&\tilde\gamma^{}_{{\rm II},k}\end{pmatrix} \end{matrix},
\end{equation}
such that $\gamma_{i,k}\left|\Psi_{\scriptscriptstyle{\rm HF}}\right>=\tilde\gamma_{i,k}\left|\Psi_{\scriptscriptstyle{\rm HF}}\right>=0$, for $i={\rm I},{\rm II}$.
Note that $\left|\Psi_{\scriptscriptstyle{\rm HF}}\right>$ depends on the effective parameters through the matrix $\mathcal{U}_k$.

We now calculate the expectation value of the full Hamiltonain $\langle H \rangle_{\scriptscriptstyle{\rm HF}}\equiv\left<\Psi_{\scriptscriptstyle{\rm HF}}\left|H\right|\Psi_{\scriptscriptstyle{\rm HF}}\right>$. The expectation value of the free Hamiltonian $H_0$ is given by
\begin{equation}
\begin{split}
\langle H_0 \rangle_{\scriptscriptstyle{\rm HF}}&=\displaystyle\sum_k \displaystyle\sum_{i,j=1}^4 \bar{\mathcal{H}}_{0,k,i,j}\langle\Phi^\dag_{k,i}\Phi_{k,j}\rangle_{\scriptscriptstyle{\rm HF}}=\displaystyle\sum_k \displaystyle\sum_{i,j,m,n=1}^4 \bar{\mathcal{H}}_{0,k,i,j}\mathcal{U}^\dag_{k,mi}\mathcal{U}_{k,jn}\langle\Gamma^\dag_{k,m}\Gamma_{k,n}\rangle_{\scriptscriptstyle{\rm HF}}=\\
&=\displaystyle\sum_k \displaystyle\sum_{m=3}^4\displaystyle\sum_{i,j=1}^4 \bar{\mathcal{H}}_{0,k,i,j}\mathcal{U}^\dag_{mi}\mathcal{U}_{jm}=\displaystyle\sum_k \Tr_{\scriptscriptstyle (-)}\left(\mathcal{U}^\dag_k\bar{\mathcal{H}}_{0,k}\mathcal{U}_k\right)
\end{split},\label{E0}
\end{equation}
where $\Tr_{\scriptscriptstyle (-)}$ is defined as a partial trace which runs only over the negative energies, i.e. only on $i=3,4$. Notice that we have omitted $E_{\rm c}$ since it does not depend on the effective parameters and therefore has no effect on the minimization procedure.
For the expectation value of the interaction Hamiltonian $H_{\rm I}$ we use Wick's theorem to arrive at
\begin{equation}
\begin{split}
\langle H_{\rm I} \rangle_{\scriptscriptstyle{\rm HF}}&=\frac{1}{L}\displaystyle\sum_{k,p,q,\sigma} U_\sigma \langle c^\dag_{\sigma,k+q\uparrow}c^\dag_{\sigma,p-q\downarrow}c_{\sigma,p\downarrow} c_{\sigma,k\uparrow}\rangle_{_{\rm HF}}=\frac{1}{L}\displaystyle\sum_{\sigma,k,p,q}U_\sigma\left[\langle c^\dag_{\sigma,k+q\uparrow}c_{\sigma,k\uparrow}\rangle_{_{\rm HF}} \langle c^\dag_{\sigma,p-q\downarrow}c_{\sigma,p\downarrow}\rangle_{_{\rm HF}}- \right.\\
&\left. - \langle c^\dag_{\sigma,k+q\uparrow}c_{\sigma,p\downarrow}\rangle_{_{\rm HF}}\langle c^\dag_{\sigma,p-q\downarrow}c_{\sigma,k\uparrow}\rangle_{_{\rm HF}}+\langle c^\dag_{\sigma,k+q\uparrow}c^\dag_{\sigma,p-q\downarrow}\rangle_{_{\rm HF}}\langle c_{\sigma,p\downarrow}c_{\sigma,k\uparrow}\rangle_{_{\rm HF}}\right]
\end{split}
\end{equation}
The Hamiltonian $H^{\scriptscriptstyle{\rm HF}}$ conserves $s^z$ and is invariant under lattice translations and therefore
\begin{equation}
\begin{array}{ccccc}
\langle c^\dag_{\sigma,k+q\uparrow}c_{\sigma,p\downarrow}\rangle_{_{\rm HF}} = 0 &,& \langle c^\dag_{\sigma,k+q,s}c_{\sigma,k,s}\rangle_{_{\rm HF}}=\langle c^\dag_{\sigma,k,s}c_{\sigma,k,s}\rangle_{_{\rm HF}}\delta_{q,0} &,&  \langle c_{\sigma,p\downarrow}c_{\sigma,k\uparrow}\rangle_{_{\rm HF}}=\langle c_{\sigma,-k\downarrow}c_{\sigma,k\uparrow}\rangle_{_{\rm HF}}\delta_{p,-k} .
\end{array} \label{eq:trans_symm}
\end{equation}
We are left with
\begin{equation}
\begin{split}
&\langle H_{\rm I}\rangle_{_{\rm HF}}=\frac{1}{L}\sum_\sigma U_\sigma\left(N_{\sigma,\uparrow}N_{\sigma,\downarrow}+\left|P_\sigma\right|^2\right)\\
&\begin{array}{c c}
N_{\sigma,s}=\sum_k \langle c_{\sigma,k,s}^\dag c_{\sigma,k,s}\rangle_{_{\rm HF}} , & P_\sigma=\sum_k \langle c_{\sigma,k,\uparrow}^\dag c_{\sigma,-k\downarrow}^\dag\rangle_{_{\rm HF}}
\end{array}
\end{split}\label{E_I}
\end{equation}
For the spin-up density terms $N_{\sigma,\uparrow}$ one has
\begin{equation}
\begin{split}
N_{a,\uparrow}&=\displaystyle\sum_k \langle c_{a,k,\uparrow}^\dag c_{a,k,\uparrow}\rangle_{_{\rm HF}}= \displaystyle\sum_k \langle \Phi_{k,1}^\dag\Phi_{k,1}\rangle_{_{\rm HF}}=\displaystyle\sum_{k,m,n} \mathcal{U}^\dag_{k,m1}\mathcal{U}_{k,1n}\langle \Gamma_{k,m}^\dag\Gamma_{k,n}\rangle_{_{\rm HF}}\\
&=\displaystyle\sum_k\displaystyle\sum_{m=3,4} \mathcal{U}^\dag_{k,m1}\mathcal{U}_{k,1m}=\displaystyle\sum_k\left(\left|\mathcal{U}_{k,13}\right|^2+ \left|\mathcal{U}_{k,14}\right|^2\right) ,
\end{split}\label{N_a_up}
\end{equation}
and similarly
\begin{equation}
N_{b,\uparrow}=\displaystyle\sum_k\left(\left|\mathcal{U}_{k,23}\right|^2+\left|\mathcal{U}_{k,24}\right|^2\right) ,\label{N_b_up}
\end{equation}
whereas for the spin-down density terms $N_{\sigma,\downarrow}$ we get
\begin{equation}
\begin{split}
N_{a,\downarrow}&=\displaystyle\sum_k \langle c_{a,k,\uparrow}^\dag c_{a,k,\uparrow}\rangle_{_{\rm HF}}= \displaystyle\sum_k \langle \Phi_{-k,3}\Phi^\dag_{-k,3}\rangle_{_{\rm HF}}=\displaystyle\sum_{k}\left(1- \displaystyle\sum_{m,n}\mathcal{U}^\dag_{k,m3}\mathcal{U}_{k,3n}\langle \Gamma^\dag_{-k,m}\Gamma_{-k,n}\rangle_{_{\rm HF}}\right)\\
&=\displaystyle\sum_k\displaystyle\left(1- \sum_{m=3,4}\mathcal{U}^\dag_{k,m3}\mathcal{U}_{k,3m}\right)=\displaystyle\sum_k\left(1-\left|\mathcal{U}_{k,33}\right|^2- \left|\mathcal{U}_{k,34}\right|^2\right)
\end{split}\label{N_a_down}
\end{equation}
and
\begin{equation}
N_{b,\downarrow}=\displaystyle\sum_k\left(1-\left|\mathcal{U}_{k,43}\right|^2- \left|\mathcal{U}_{k,44}\right|^2\right) .\label{N_b_down}
\end{equation}
Using the same manipulations for the pairing terms $P_\sigma$, we obtain
\begin{equation}
\begin{array}{c c c}
P_a=\displaystyle\sum_k\mathcal{U}^*_{k,1,3}\mathcal{U}_{k,33}+\mathcal{U}^*_{k,14}\mathcal{U}_{k,34}&, & P_b=\displaystyle\sum_k\mathcal{U}^*_{k,2,3}\mathcal{U}_{k,43}+\mathcal{U}^*_{k,24}\mathcal{U}_{k,44}.
\end{array}\label{P_a_b}
\end{equation}
Finally, we use Eqs.~(\ref{E0},~\ref{E_I}-\ref{P_a_b}) to calculate $\langle H\rangle_{_{\rm HF}}$, and numerically minimize it with respect to the effective parameters $\tilde\mu_a, \tilde\mu_b,\tilde\Delta_{\rm ind}$ and $\tilde\Delta_b$.

\subsection{Local Stability to Spin-Density Waves}\label{SDW}
So far in our Hartree-Fock treatment, we did not account for the possibility that the system spontaneously breaks the lattice translational symmetry, and develops spin-density waves. To examine this possibility, one should consider the following more general Hartree-Fock Hamiltonian:
\begin{equation}
H^{\scriptscriptstyle{\rm HF}}(\phi) = H^{\scriptscriptstyle{\rm HF}}(0)-\sum_{\sigma,q,s}\phi_{\sigma -q s}\hat{\rho}_{\sigma q s} ,
\end{equation}
where $\hat{\rho}_{\sigma q s}$ is the Fourier transform of $\hat{n}_{\sigma i s}=c^\dag_{\sigma i s}c_{\sigma i s}$, given by $\hat{\rho}_{\sigma q s}=\frac{1}{\sqrt{L}}\sum_k c^\dag_{\sigma k+q s}c_{\sigma k s}$, $\phi_{\sigma q s}$ are fields which in general should be determined by the Hartree-Fock procedure, and $H^{\scriptscriptstyle{\rm HF}}(0)$ is the same as the Hamiltonian of Eq.~\eqref{eq:HF_Hamiltonian}. We now ask whether the addition of the fields $\phi_{\sigma q s}$ can cause the expectation value of the full Hamiltonian $\langle H\rangle_{{\scriptscriptstyle{\rm HF}},\phi}\equiv\left<\Psi_{\scriptscriptstyle{\rm HF}}(\phi)\left|H\right|\Psi_{\scriptscriptstyle{\rm HF}}(\phi)\right>$ to decrease, assuming that these fields are small. If this is the case then our previous Hartree-Fock solution is locally unstable to the formation of spin-density waves. In other words, for the solution to be locally stable, $\langle H\rangle_{{\scriptscriptstyle{\rm HF}},\phi}$ must have a minimum at $\phi_{\sigma,q,s}=0$. This is the case if the Hessian matrix $\left.\partial^2{\langle H\rangle_{_{\rm HF},\phi}}/\partial\phi_{\sigma',q,s'}\partial\phi_{\sigma,-q,s}\right|_{\phi=0}$ is positive definite.

We start by rewriting the expression for $\langle H\rangle_{{\scriptscriptstyle{\rm HF}},\phi}$,
\begin{equation}
\langle H\rangle_{{\scriptscriptstyle{\rm HF}},\phi}=\langle H_0\rangle_{{\scriptscriptstyle{\rm HF}},\phi} + \frac{1}{L}\sum_\sigma U_\sigma\left(N_{\sigma,\uparrow}N_{\sigma,\downarrow}+\left|P_\sigma\right|^2\right)+ \displaystyle\sum_\sigma U_\sigma\displaystyle\sum_{q\neq0}\rho_{\sigma,q\uparrow}\rho_{\sigma,-q\downarrow} ,\label{eq:H_avg}
\end{equation}
where $\rho_{\sigma,qs}\equiv\langle\hat{\rho}_{\sigma,qs}\rangle_{_{\rm HF},\phi}$. We now write $H_0$ as
\begin{equation}
H_0=H^{\scriptscriptstyle{\rm HF}}(\phi)+ \sum_\sigma \delta\mu_\sigma\sum_{k,s} c^\dag_{\sigma,k,s}c_{\sigma,k,s} - \sum_\sigma \delta\Delta_\sigma\sum_k \left(c^\dag_{\sigma,k\uparrow}c^\dag_{\sigma,-k\downarrow}+h.c.\right) +\sum_{\sigma,q\neq0,s}\phi_{\sigma -q s}\hat{\rho}_{\sigma q s},
\end{equation}
where we have defined $\delta\mu_\sigma\equiv\tilde{\mu}_\sigma-\mu_\sigma$, $\delta\Delta_a\equiv\tilde{\Delta}_{\rm ind}-\Delta_{\rm ind}$, and $\delta\Delta_b\equiv\tilde{\Delta}_b-\Delta_b$. Inserting this into Eq.~\eqref{eq:H_avg}, one has
\begin{equation}
\begin{split}
\langle H\rangle_{{\scriptscriptstyle{\rm HF}},\phi}&=\langle H^{\scriptscriptstyle{\rm HF}}(\phi)\rangle_{{\scriptscriptstyle{\rm HF}},\phi}+ \sum_{\sigma}N_\sigma \left(\frac{U_\sigma}{L}N_\sigma+2\delta\mu_\sigma\right) + \sum_\sigma P_\sigma \left(\frac{U_\sigma}{L}P_\sigma-2\delta\Delta_\sigma\right) +\\ & + \displaystyle\sum_{\sigma,q\neq0} \left(U_\sigma\rho_{\sigma,-q\downarrow}\rho_{\sigma,q\uparrow} +\sum_s \phi_{\sigma -q s}\rho_{\sigma qs}\right).
\end{split}\label{eq:H_avg_2}
\end{equation}
In the last step we have used the fact that due to TR symmetry, $N_{\sigma,\uparrow}=N_{\sigma,\downarrow}\equiv N_\sigma$ and $P^*_\sigma=P_\sigma$.
To perform the derivative of the first term, we exploit the Hellmann-Feynman theorem
\begin{equation}
\frac{\partial\langle H^{\scriptscriptstyle{\rm HF}}(\phi)\rangle_{{\scriptscriptstyle{\rm HF}},\phi}}{\partial \phi_{\sigma,q,s}}=\left<\frac{\partial H^{\scriptscriptstyle{\rm HF}}(\phi)}{\partial \phi_{\sigma,q,s}}\right>_{{\scriptscriptstyle{\rm HF}},\phi}=-\rho_{\sigma,-q,s}
\end{equation}
We now invoke linear response theory:
\begin{equation}
\rho_{\sigma,q,s} = \sum_{\sigma'}\chi_{q,\sigma,s}^{\sigma'}\phi_{q,\sigma',s}.
\end{equation}
Note that due to $s^z$ conservation, there is no response of $\rho_{\sigma,q,\uparrow}$ to $\phi_{q,\sigma',\downarrow}$ and vice versa. We next perform the second derivative of $\langle H\rangle_{{\scriptscriptstyle{\rm HF}},\phi}$, and we note that the second and third terms in Eq.~\eqref{eq:H_avg_2} then vanish as a result of the self-consistent equations for $\delta\mu_\sigma$ and $\delta\Delta_\sigma$. One then obtains $\langle H\rangle_{{\scriptscriptstyle{\rm HF}},\phi}$ to second order in $\phi$ using the Hessian matrix:
\begin{equation}
\begin{array}{c c c}
\langle H\rangle_{{\scriptscriptstyle{\rm HF}},\phi}=\frac{1}{2}\displaystyle\sum_{q\neq0}\vec{\phi}^T_{-q} h_q \vec{\phi}_q & ,&
\vec{\phi}^T_{q}=\begin{pmatrix} \phi_{qa\uparrow} & \phi_{qb\uparrow} & \phi_{qa\downarrow} & \phi_{qb\downarrow}\end{pmatrix}
\end{array}\label{eq:Hessian_1}
\end{equation}

\begin{equation}
\begin{array}{c c c c c}
h_q=\begin{pmatrix} A_q&B_q\\ B_q&A_q\end{pmatrix} &,&
A_q=\begin{pmatrix}\chi_{aa,q}&\chi_{ab,q}^*\\ \chi_{ab,q} & \chi_{bb,q}\end{pmatrix} &, &
B_q=\begin{pmatrix}U_a\chi_{aa,q}^2+U_b|\chi_{ab,q}|^2 &\chi_{ab,q}^*\left(U_a\chi_{aa,q}+U_b\chi_{bb,q}\right)\\ \chi_{ab,q}\left(U_a\chi_{aa,q}+U_b\chi_{bb,q}\right) & U_a|\chi_{ab,q}|^2+U_b\chi_{bb,q}^2 \end{pmatrix}
\end{array}.\label{eq:Hessian_2}
\end{equation}
In writing Eq.~\eqref{eq:Hessian_1} and Eq.~\eqref{eq:Hessian_2}, we made use of the fact that due to TR symmetry $\chi_{\sigma,q,\uparrow}^{\sigma'}= \chi_{\sigma',-q,\downarrow}^{\sigma}\equiv \chi_{\sigma,\sigma',q}$, as well as of $\chi^*_{\sigma,\sigma',-q} =\chi_{\sigma,\sigma',q}$.

We calculate the susceptibilities using the Kubo Formula:
\begin{equation}
\chi^{\sigma'}_{\sigma,q,s}=-i\int_{-\infty}^\infty{\theta(t)\left<\left[\hat{\rho}_{-q,\sigma,s}(0),\hat{\rho}_{q,\sigma',s}(t)\right]\right>_{_{\rm HF}} dt} ,
\end{equation}
and obtain
\begin{equation}
\begin{split}
\chi_{aa,q}&=\frac{1}{L}\displaystyle\sum_k\sum_{i,j=1}^2 \frac{|u_{1i}(k)|^2|v_{1j}(k+q)|^2 + |u_{1j}(k+q)|^2|v_{1i}(k)|^2} {\epsilon_{k,i}+\epsilon_{k+q,j}} \\
\chi_{bb,q}&=\frac{1}{L}\displaystyle\sum_k\sum_{i,j=1}^2 \frac{|u_{2i}(k)|^2|v_{2j}(k+q)|^2 + |u_{2j}(k+q)|^2|v_{2i}(k)|^2} {\epsilon_{k,i}+\epsilon_{k+q,j}}\\
\chi_{ab,q}&=\frac{1}{L}\displaystyle\sum_k\sum_{i,j=1}^2 \frac{u_{1i}(k)u^*_{2i}(k)v^*_{1j}(k+q)v_{2j}(k+q)+ u^*_{1j}(k+q)u_{2j}(k+q)v_{1i}(k)v^*_{2i}(k)} {\epsilon_{k,i}+\epsilon_{k+q,j}} .
\end{split}\label{eq:susceptibility}
\end{equation}
Finally, we use Eqs.~\eqref{eq:susceptibility} to numerically check that $h_q$ in Eq.~\eqref{eq:Hessian_2} is positive definite (i.e. that all its eigen-values are positive) for all $q$'s as a condition for local stability.

\subsection{The DIII Topological Invariant}

We hereby show a method for the calculation of the $\mathbb{Z}_2$ topological invariant of the DIII symmetry class, and apply it to the Hartree-Fock Hamiltonian.
Let the Hamiltonian be written as
\begin{equation}
\begin{array}{c c c}
H=\displaystyle\sum_k\Psi^\dag_k\mathcal{H}_k\Psi_k&,&
\Psi^\dag_k=\begin{pmatrix}\psi^\dag_k&,-is^y\psi_k\end{pmatrix}
\end{array}.
\end{equation}
Time-reversal and particle-hole symmetries are given in this basis by
\begin{equation}
\begin{array}{ccc}
\mathcal{T}=s^yK &,& \mathcal{T}\mathcal{H}_k\mathcal{T}^{-1}=\mathcal{H}_{-k}\\ \mathcal{P}=s^y\tau^yK &,& \mathcal{P}\mathcal{H}_k\mathcal{P}^{-1}=-\mathcal{H}_{-k}\\
\end{array},
\end{equation}
where $K$ stands for complex conjugation. We therefore also have a chiral symmetry expressed by
\begin{equation}
\begin{array}{ccc}
\mathcal{C}=\tau^y &,& \left\{\mathcal{C},\mathcal{H}_k\right\}=0
\end{array}.
\end{equation}
The unitary transformation $\mathcal{V}=e^{i\frac{\pi}{4}\tau^x}$ diagonalizes $\mathcal{C}$ and brings the Hamiltonian to the following form
\begin{equation}
\tilde{\mathcal{H}}_k=\mathcal{V}\mathcal{H}_k\mathcal{V}^\dag= \begin{pmatrix}0&\mathcal{B}_k\\ \mathcal{B}^\dag_k&0\end{pmatrix}.
\end{equation}
Keeping the system gapped, we can deform $\mathcal{B}_k$ to be unitary by adiabatically taking all its singular values to unity. This is equivalent to deforming the spectrum of $\tilde{\mathcal{H}}_k$ into two flat bands with energies $\pm E$, without changing the eigenfunctions. The time-reversal operator is given in this basis by
\begin{equation}
\tilde{\mathcal{T}}=\mathcal{V}\mathcal{T}\mathcal{V}^\dag=is^y\tau^xK,
\end{equation}
allowing us to express the time-reversal symmetry in terms of $\mathcal{B}_k$ as
\begin{equation}
s^y\mathcal{B}^\dag_ks^y=\mathcal{B}^*_{-k}.
\end{equation}
We can now formulate Kramers' theorem for the matrix $\mathcal{B}_k$: if $w_k$ is an eigen-vector of $\mathcal{B}_k$ with an eigen-value $e^{i\theta_k}$, then $s^yw^*_k$ is an eigen-vector of $\mathcal{B}_{-k}$ with the same eigen-value. In particular, at the TR invariant points $k=0,\pi$, one has a protected Kramers' degeneracy. We next divide the eigen-values of $\mathcal{B}_k$ to $\{e^{i\theta_{n,k}}\}_n$ and their time-reversed partners $\{e^{i\tilde{\theta}_{n,k}}\}_n$, where $\tilde{\theta}_{n,-k}=\theta_{n,k}$, and sum over the winding numbers of the former only:
\begin{equation}
N=\frac{1}{2\pi}\displaystyle\sum_n\int_{k=-\pi}^{k=\pi} d\theta_{n,k}.
\end{equation}
The parity of this winding number $\nu=\mod(N,2)$ is a topological invariant.

Applying the above procedure to the $8\times8$ Hartree-Fock Hamiltonian $\mathcal{H}^{\scriptscriptstyle{\rm HF}}_k$, described in the main text, one has that
\begin{equation}
\begin{array}{ccc}
\mathcal{B}_k=\begin{pmatrix} \mathcal{B}^\uparrow_k& \\ & \mathcal{B}^\downarrow_k\end{pmatrix} &,&
\mathcal{B}^\uparrow_k=\mathcal{B}^\downarrow_{-k}=\begin{pmatrix} \tilde\Delta_{\rm ind}-i\tilde\varepsilon_{a,k} & -it_{ab}\\-it_{ab}& \tilde\Delta_b-i\tilde\varepsilon_{b,k}\end{pmatrix}
\end{array}.
\end{equation}
The $s^z$ symmetry of $\mathcal{H}^{\scriptscriptstyle{\rm HF}}_k$ thus allows us to easily separate the eigen-values $\{e^{i\theta_{n,k}}\}_n$ from their time-reversed partners. The topological invariant is then simply given as the parity of the winding number of
\begin{equation}
\det{\mathcal{\mathcal{B}}^\uparrow_k} = t_{ab}^2+\tilde\Delta_{\rm ind}\tilde\Delta_b -\tilde\varepsilon_{a,k}\tilde\varepsilon_{b,k}-i(\tilde\Delta_{\rm ind}\tilde\varepsilon_{b,k}+\tilde\Delta_b\tilde\varepsilon_{a,k}) .
\end{equation}

\subsection{DMRG}

\begin{figure}[h]
\subfloat[]{\includegraphics[width=0.4\textwidth]{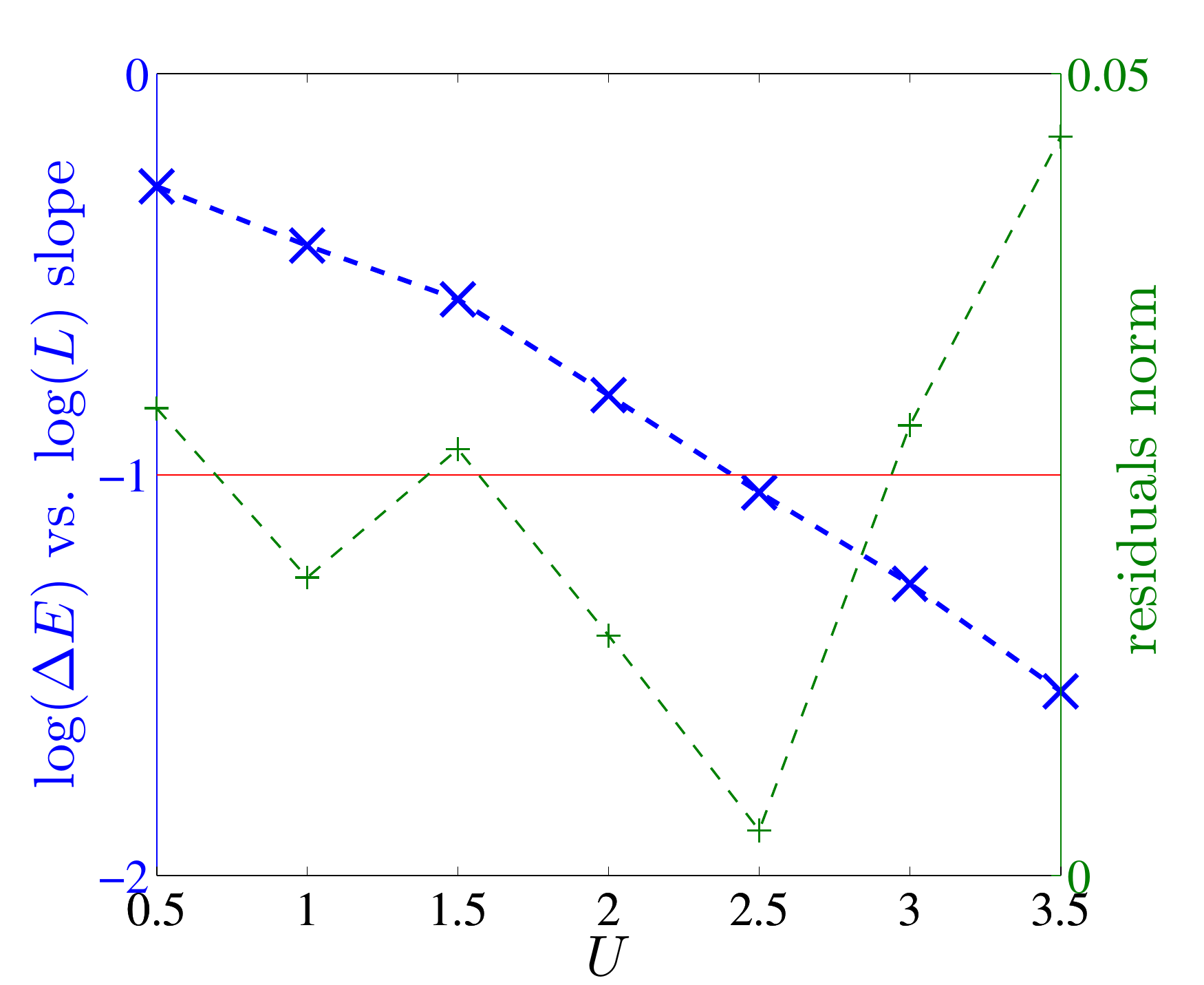}}
\subfloat[]{\includegraphics[width=0.4\textwidth]{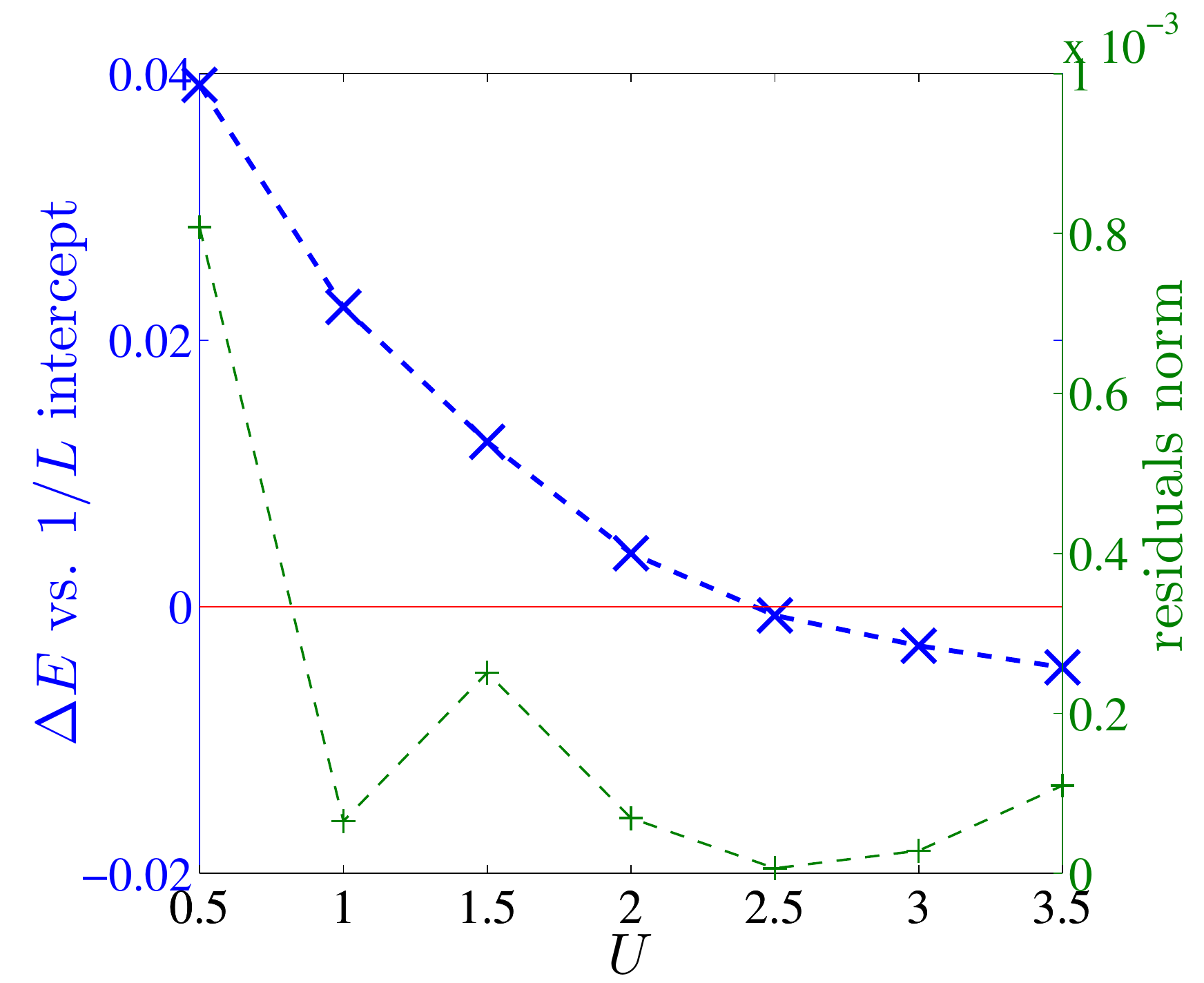}}
\caption{\label{fig:PhaseTransition}Analysis of the scaling of the gap between
the ground state and the first excited state with system size, as
the on-site repulsive interaction strength is varied. The system's
parameters are identical to the ones in Fig. 3 of the main text: $t_{a}=t_{b}=1,\ t_{ab}=0.4,\ \alpha_{a}=0,\ \alpha_{b}=0.6,\ \Delta_{{\rm ind}}=1,\ \mu=0.8$.
(a) The blue line is the $\log\left(\Delta E\right)$-$\log(L)$ slope.
The system starts in a trivial superconducting phase for small enough
$U$'s with slope close to zero. As the slope reaches $-1$ (indicated
by the red line) at $U=U_{c}\simeq2.4$, the system undergoes a phase
transition and continues to the TRITOPS phase with slopes $<-1$.
The green curve is the norm of the residuals for a linear fit. It
can be seen that the residuals are smallest close to the critical
point, indicating the agreement of the curve with a linear fit as
expected. (b) The blue line is the intercept of a linear fit to $\Delta E$-$1/L$
plot with the $\Delta E$ axis. The intercept point is first positive,
suggesting the system is in the trivial superconducting phase, then
crosses zero (indicated by the red line) at $U=U_{c}\simeq2.4$ as
a phase transition occurs, and continues to be negative in agreement
with our expectations for the TRITOPS phase. The green curve is once
again the norm of the residuals for a linear fit. As expected, the
residuals are smallest close to the critical point.}
\end{figure}

To obtain the DMRG phase diagram presented in the main text we analyze
the scaling of the low-lying energy spectrum with the length of the wire, $L$.
In the trivial superconducting phase the system is gapped. In the limit $L\rightarrow\infty$ we denote this gap by $\Delta$.
The first excited state in this system can be approximately described using a single particle picture with a parabolic dispersion above the gap. We therefore expect the finite size correction to the excitation energy to be quadratic in $1/L$, i.e. $\Delta E_{\rm Trivial}=\Delta+C\left(\frac{1}{L}\right)^{2}$.
The low-energy effective theory at the critical point is given by a massless Dirac fermionic mode, that becomes a Luttinger liquid in the presence of interactions.
Hence, in a finite wire we expect the excitation energies to be inversely proportional to the system size, $\Delta E_{\rm Critical}\propto\frac{1}{L}$.
In the TRITOPS phase, the ground state is fourfold degenerate in the thermodynamic limit. In
a finite system we expect exponential splittings between the states
within the ground state manifold. In particular, the splitting between
the two lowest energy states is $\Delta E_{\rm TRITOPS}\propto e^{-L/\xi}$,
where $\xi$ is the coherence length of the superconductor.
We first discuss two ways to determine the phase transition point
and then implement them for our system.

\medskip
\emph{Method I - Slope of $log\left(\Delta E\right)$ vs. $log(L)$}

Consider first the slope of the curve given by $\log\left(\Delta E\right)$
vs. $\log(L)$. For a trivial superconductor $\log\left(\Delta E_{\rm Trivial}\right)\rightarrow \log\left(\Delta\right)={\rm const.}$
as $L\rightarrow\infty$, and the slope tends to zero. At the critical
point $\log\left(\Delta E_{\rm Critical}\right)\propto-\log\left(L\right)$
and we therefore expect the curve to be linear with a slope equal
to $-1$. For a system in the TRITOPS phase, $\log\left(\Delta E_{\rm TRITOPS}\right)=-\frac{L}{\xi}+{\rm const.}=-\frac{1}{\xi}e^{\log\left(L\right)}+{\rm const.}$.
The slope of the $\log-\log$ curve is then $-\frac{L}{\xi}$. For a large
enough system (larger than the coherence length) the slope is less
than $-1$. Therefore, as we go from a trivial superconducting phase
through a critical point into the TRITOPS phase, we expect the slope
to start from values close to zero, pass through $-1$ at the phase
transition and continue to be less than $-1$.

\medskip
\emph{Method II - Intercept of $\Delta E$ vs. $1/L$ with $\Delta E$ axis}

Consider next the energy gap plotted on $\Delta E$ vs. $1/L$ axes.
One can then calculate the intercept of a linear fit to the data with
the $\Delta E$ axis. For a trivial superconductor $\Delta E\rightarrow\Delta={\rm const.}>0$
as $\frac{1}{L}\rightarrow0$, therefore for large enough systems
the intercept point will be positive. At the critical point $\Delta E$
is exactly proportional to $\frac{1}{L}$, hence the linear fit should
be asymptotically exact with the intercept at zero. In the TRITOPS phase $\Delta E\rightarrow0$
as $\frac{1}{L}\rightarrow0$, but the curvature is positive for large
enough but finite $L$'s. Therefore, the intercept of a linear fit
with the $\Delta E$ axis will be negative.
We therefore expect that as we go from a trivial superconducting phase
through a critical point into the TRITOPS phase, the intercept will
start positive, pass through $0$ at the phase transition and continue
to be negative.

\medskip
\emph{The phase diagram}

We now scan the on-site repulsive interactions strength $U$ at fixed $\mu$, and use
the above analysis to find the phase transition point.
Note that in all the three cases the ground state energy is the lowest
energy in the even fermion parity sector, while the first excited
state is the lowest energy state in the odd fermion parity sector.
It is therefore sufficient to target a single state in each fermion
parity sector to obtain the energy gap.
For each interaction strength $U$ we calculate the energy gap for
three different lengths of the wire $L=60,100,140$. We then plot
the data on $\log\left(\Delta E\right)$ vs. $\log(L)$ axes and calculate
the slope, as well as on $\Delta E$ vs. $1/L$ axes and calculate
the intercept point of a linear fit to the data with the $\Delta E$
axis.
The analysis using both methods is shown in Fig. \ref{fig:PhaseTransition}.
It can be seen that both methods agree and suggest a phase transition
from a trivial superconducting phase to a TRITOPS phase occurs at
$U_{c}\simeq2.4$.

\medskip
In all DMRG calculations presented in this work we 
kept up to 800 states, and the typical truncation errors were of order $10^{-5}$.

\subsection{Additional Results}
\label{Additional_results}
\emph{Hartree-Fock.}$-$Below we present Hartree-Fock results for different sets of parameters than those given in the main text. Fig.~\ref{fig:HF_phase_diagram_diff_params} shows the Hartree-Fock phase diagram for the set of parameters $t_a=t_b=1, t_{ab}=0.4, \alpha_a=0, \alpha_b=1$ and $\Delta_{\rm ind}=1$, while Fig.~\ref{fig:HF_phase_diagram_diff_params_2} shows results for $t_a=t_b=1, t_{ab}=1, \alpha_a=-0.6, \alpha_b=0.6$ and $\Delta_{\rm ind}=1$.

\begin{figure}[h]
\subfloat[]{\includegraphics[clip=true,trim =2.5cm 8.5cm 1.5cm 9cm,width=0.5\textwidth]{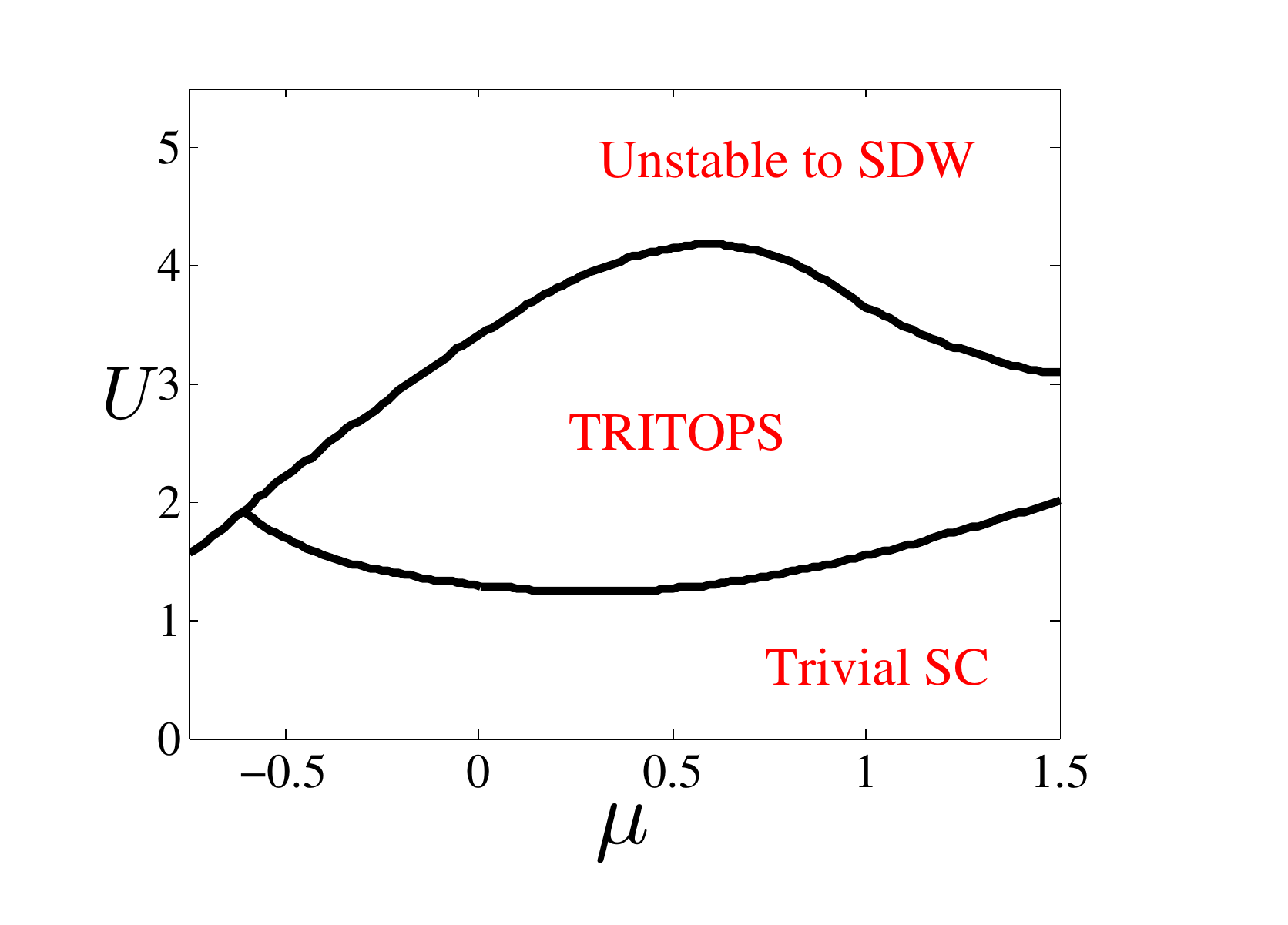}\label{fig:HF_phase_diagram_diff_params}}
\subfloat[]{\includegraphics[clip=true,trim =2.5cm 8.5cm 1.5cm 9cm,width=0.5\textwidth]{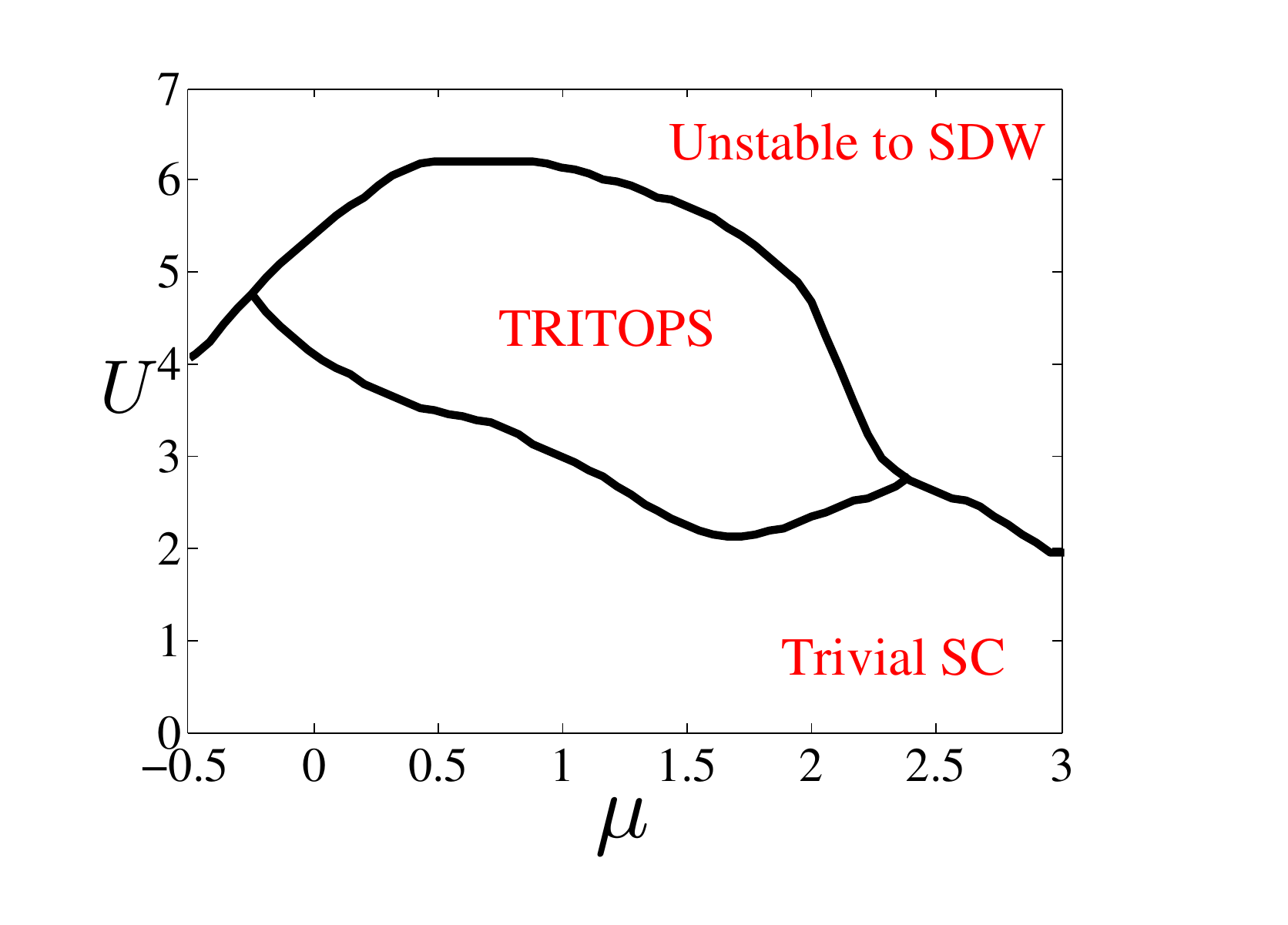}\label{fig:HF_phase_diagram_diff_params_2}}
\caption{Hartree-Fock phase diagram as a function of chemical potential $\mu_a=\mu_b=\mu$, and interaction strength $U$, for (a) $t_a=t_b=1, t_{ab}=0.4, \alpha_a=0, \alpha_b=1$
, and $\Delta_{\rm ind}=1$, and for (b) $t_a=t_b=1, t_{ab}=1, \alpha_a=-0.6, \alpha_b=0.6$
, and $\Delta_{\rm ind}=1$. The diagram includes a TR-invariant topological superconductor phase (TRITOPS), a trivial superconductor phase, and a region in which the Hartree-Fock solution is locally unstable to the formation of spin-density waves.}\label{fig:HF_phase_diagram_SM}
\end{figure}

\medskip
\emph{DMRG.}$-$In addition, we present the low lying energy spectrum obtained using DMRG for the same set of parameters used in the main text
($t_{a}=t_{b}=1,\ t_{ab}=0.4,\ \alpha_{a}=0,\ \alpha_{b}=0.6,\ \Delta_{{\rm ind}}=1,\ \mu=0.8$),
but for larger repulsive interactions strength $U=8>U_c$ in Fig~\ref{fig:EnTopologicalU8}.
It can be clearly seen that the system is still in the topological phase with a finite gap.
Note that the gap extracted by extrapolating $1/L\rightarrow0$ for $U=8$ is $\Delta E\simeq0.02$,
i.e. reduced only by about a factor of $2$ with respect to the gap for $U=5.5$ presented in
Fig. 3d of the main text.

\begin{figure}[h]
\includegraphics[width=0.4\textwidth]{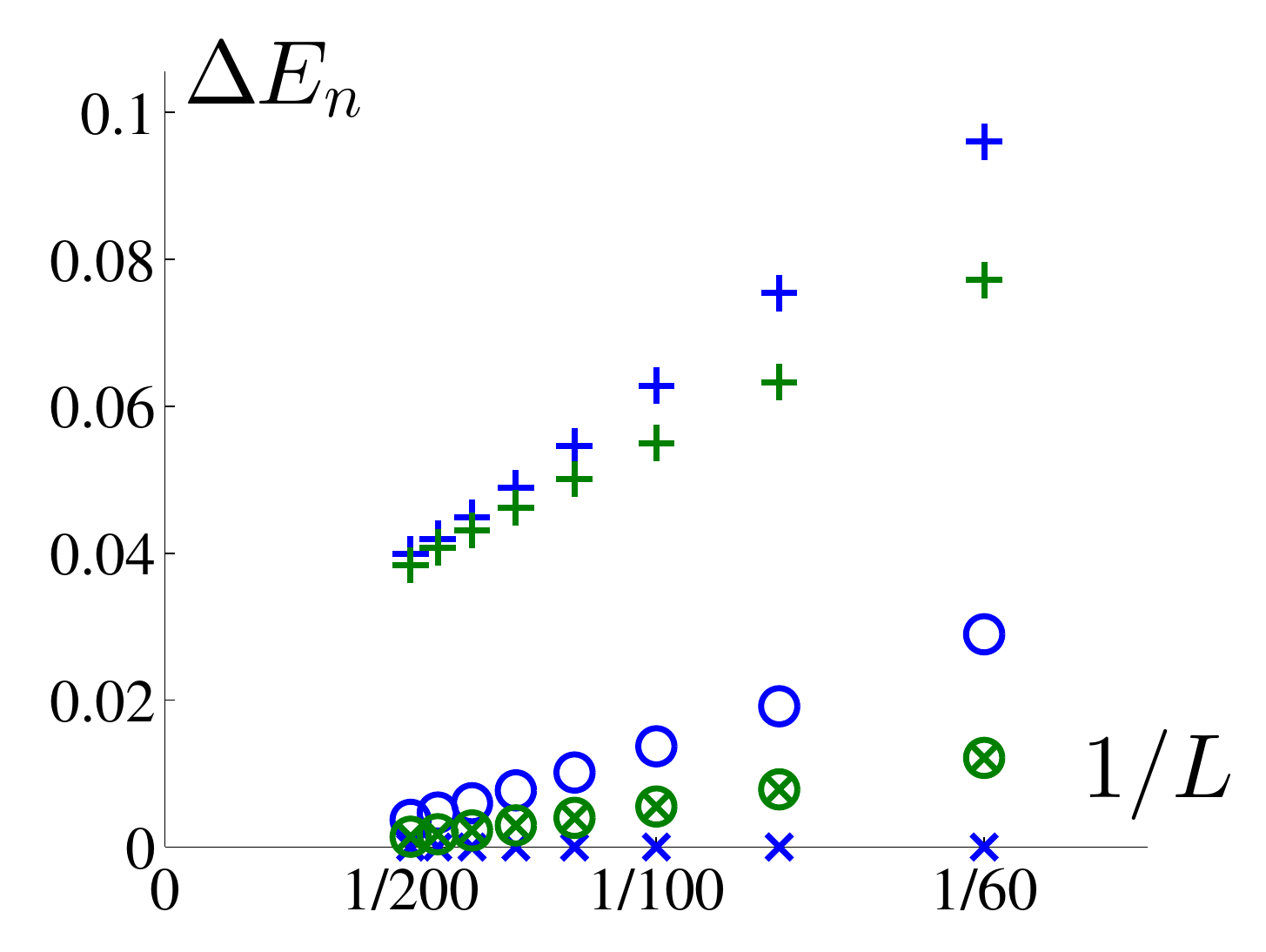}
\caption{\label{fig:EnTopologicalU8} The low-lying energy spectrum of the system vs. $1/L$, where
$L$ is the length of the wire, for system's
parameters identical to the ones in Fig. 3 of the main text:
$t_{a}=t_{b}=1,\ t_{ab}=0.4,\ \alpha_{a}=0,\ \alpha_{b}=0.6,\ \Delta_{{\rm ind}}=1,\ \mu=0.8$ ,
for interaction strength $U=8>U_c$. It is clearly seen that the system is still
in the topological phase with a fourfold degenerate ground state and a finite gap in the $L\rightarrow\infty$ limit.}
\end{figure}

\subsection{Order parameter v.s. pairing potential}
When calculating the Hartree-Fock effective parameters from which one extracts the phase diagram, we have mentioned that the pairing potential on chain $b$ has an opposite sign with respect to the one on chain $a$, namely $\sgn(\tilde{\Delta}_b)=-\sgn(\tilde{\Delta}_{\rm{ind}})$. It is constructive to observe the behavior of these potentials as a function of interaction strength $U$, and to compare it to the behavior of the superconducting order parameters. As seen in Fig.~\ref{fig:pairing}, while the pairing potentials have opposite signs, the order parameters on the two chains both have the same signs. This is determined by the sign of the adjacent bulk superconductor. As expected, as the interaction strength increases the pairing potentials and the order parameters all go to zero.
\begin{figure}[h]
\subfloat[]{\includegraphics[clip=true,trim =0cm 0cm 0cm 0cm,width=0.5\textwidth]{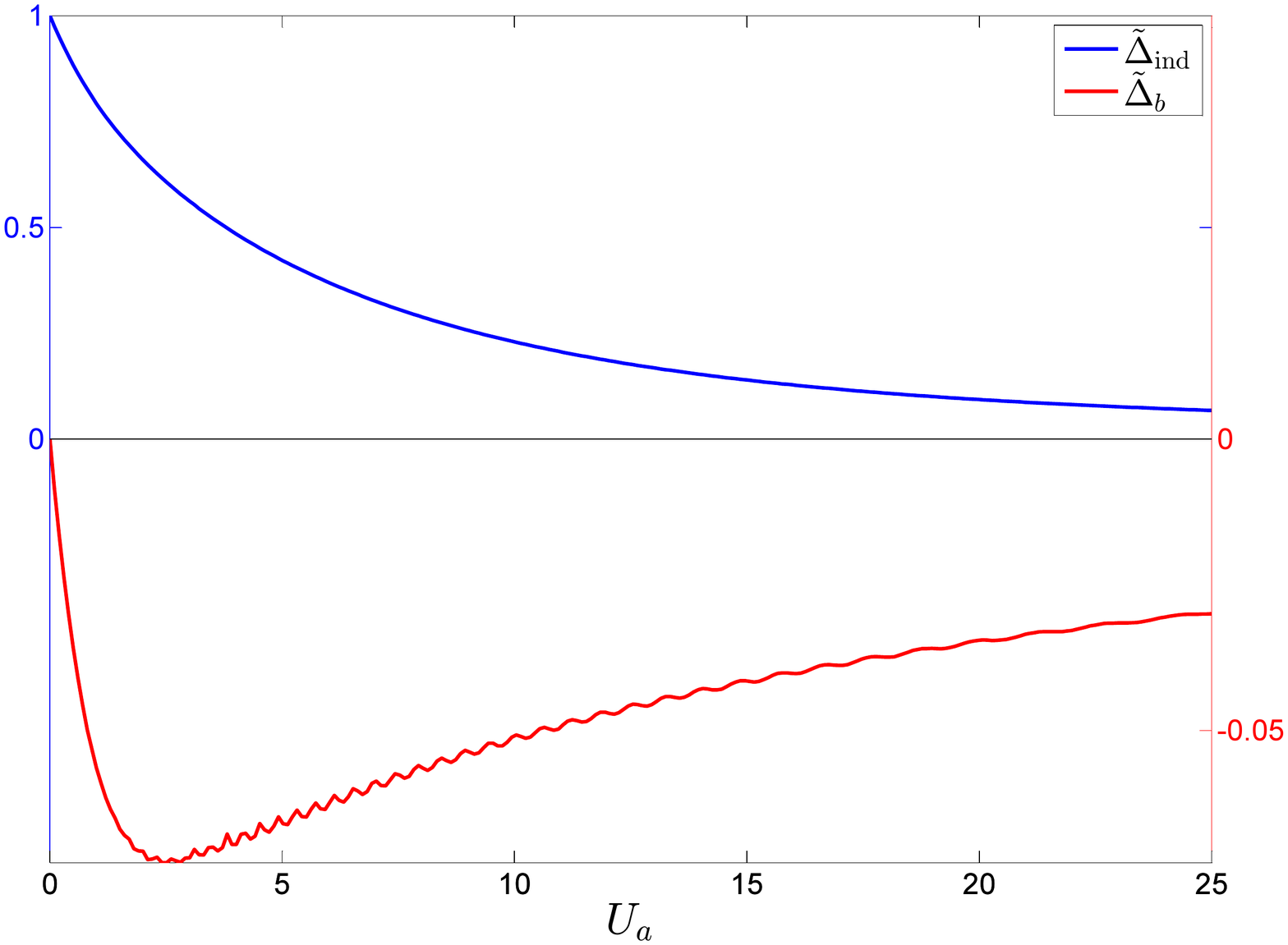}\label{fig:pairing_potential}}
\subfloat[]{\includegraphics[clip=true,trim =0cm 0cm 0cm 0cm,width=0.5\textwidth]{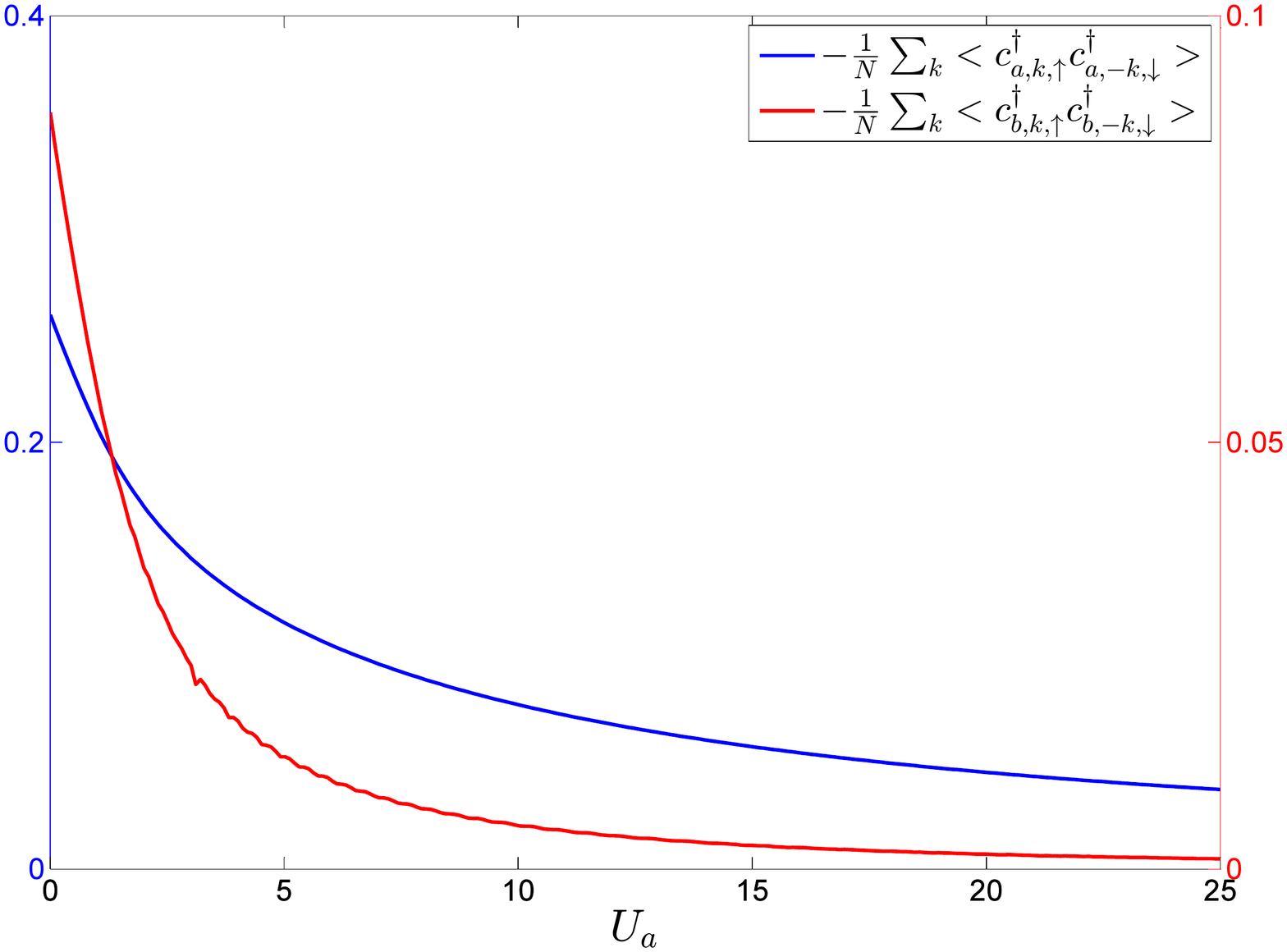}\label{fig:pairing_correlations}}
\caption{(a) Pairing potential and (b) order parameter on the two chains as a function of interaction strength $U_a=U_b$ for the parameters used to plot Fig.~\ref{fig:HF_phase_diagram_diff_params_2} and $\mu_a=\mu_b=1$. While the pairing potentials have opposite signs with respect to the two chains, the order parameters both have the same sign (that of the adjacent bulk superconductor). Note that as $U$ increases the pairing potentials and the order parameters go to zero. The observed oscillatory behavior is due to finite size effects.}\label{fig:pairing}
\end{figure}

\subsection{Estimation of model parameters}
We wish to relate the values of the model parameters (cf. Eq.~(1) of the main text) to a realistic setup of a proximity coupled Rashba wire. There are three important energy scales which defines the problem: The induced pairing potential $\Delta_{\rm{ind}}$, the interaction strength $U$, and the spin-orbit coupling energy $E_{\rm{SO}}=\alpha^2/t$.
To estimate these energy scales we borrow from the setup used in Mourik {\it et al.}~\cite{mourik2012signatures}, namely $E_{\rm{SO}}=0.05meV$ and $\Delta_{\rm{ind}}=0.25meV$. A realistic interaction strength was estimated in the main text to be $U\sim1-10K\simeq0.1-1meV$. This yields the following ratios between the energy scales:
\begin{equation}
\begin{matrix}
\frac{E_{\rm{SO}}}{\Delta_{\rm{ind}}}=0.2&,&\frac{U}{\Delta_{\rm{ind}}}=0.4-4
\end{matrix}.\label{eq:erg_scales}
\end{equation}
This should be compared to the ratios extracted from the values of the parameters used in the present calculations. For that, we first note that the spin-orbit coupling energy is given by $E_{\rm{SO}}=\alpha^2/t$. We take $t=1$ $\alpha=0.6$ and $\Delta_{\rm{ind}}=1$ in accordance with Fig.~(2-3) of the main text. The topological region as taken from the DMRG results exists in the range $U\sim2-8$. This yields
\begin{equation}
\begin{matrix}
\frac{E_{\rm{SO}}}{\Delta_{\rm{ind}}}=0.36&,&\frac{U}{\Delta_{\rm{ind}}}=~2-8
\end{matrix},
\end{equation}
in reasonable agreement with Eq.~\eqref{eq:erg_scales}.
We can further give an estimate of the energy gap in the topological region. From the DMRG results Fig.~(3.d) of the main text we have $E_{\rm gap}=0.04\cdot\Delta_{\rm ind}=10\mu eV\simeq100mK$.

\subsection{Calculation of the BDI Topological Invariant}
As one introduces a Zeeman field perpendicular to the direction of SOC, the Hartree-Fock Hamiltonian no longer respects the TR symmetry $\Theta=is^yK$, however it possesses an effective TR symmetry $\Lambda=s^xK$, expressed by $\Lambda\mathcal{H}^{\scriptscriptstyle{\rm HF}}_{k}\Lambda^{-1}=\mathcal{H}^{\scriptscriptstyle{\rm HF}}_{-k}$. This symmetry satisfies $\Lambda^2=1$, making the Hamiltonian in the BDI symmetry class~\cite{Altland1997} with a $\mathbb{Z}-$topological invariant~\cite{schnyder2008classification,kitaev2009periodic}. This invariant counts the number of Majorana modes at each end of the system. To calculate this invariant for our system we follow the recipe given in Ref.~\cite{tewari2012topological}. In accordance with this recipe, we start by writing the Hamiltonian such that it would be real when written in real space. We therefore set the SOC to be in the $\hat{y}$ direction, and the Zeeman field to be in the $\hat{x}$ direction. Furthermore, we write the BdG Hamiltonian in the basis $\varphi^\dag_k=\begin{pmatrix}\psi^\dag_k&,\psi_{-k}\end{pmatrix}$, therefore
\begin{equation}
\begin{array}{ccc}
H=\displaystyle\sum_k\varphi^\dag_k\tilde{\mathcal{H}}_k\varphi_k&,& \tilde{\mathcal{H}}_k =\left[\boldsymbol{\xi}_k+ \boldsymbol{\alpha}_ks^y +B_xs^x\right]\tau^z+\boldsymbol{\Delta}s^y\tau^y,
\end{array}
\end{equation}
where
\begin{equation}
\begin{array}{ccccc}
\boldsymbol{\xi}_k=\begin{pmatrix}\tilde\xi_{a,k}&t_{ab} \\ t_{ab}&\tilde\xi_{b,k}\end{pmatrix}&,&\boldsymbol{\alpha}_k=\begin{pmatrix}\alpha_a\sin{k}& \\ &\alpha_b\sin{k}\end{pmatrix} &,& \boldsymbol{\Delta}=\begin{pmatrix}\tilde\Delta_a& \\ &\tilde\Delta_b\end{pmatrix}.
\end{array}
\end{equation}
In this basis the new TR symmetry is given by $\tilde\Lambda=K$, and PH symmetry is given by $\tilde\Xi=\tau^xK$. The presence of a chiral symmetry, given by $\tilde{\mathcal{C}}=\tilde\Xi\tilde\Lambda$, which anti-commutes with the Hamiltonian $\{\tilde{\mathcal{C}},\tilde{\mathcal{H}}_k\}=0$, allows us to transform $\tilde{\mathcal{H}}_k$ to the following off-diagonal form
\begin{equation}
\begin{array}{ccc}
e^{-i\frac{\pi}{4}\tau^y}\tilde{\mathcal{H}}_ke^{i\frac{\pi}{4}\tau^y}=\begin{pmatrix} &\mathcal{A}_k\\ \mathcal{A}^T_{-k}& \end{pmatrix}&,&
\mathcal{A}_k = \boldsymbol{\xi}_k+\boldsymbol{\alpha}_ks^y+Bs^x+i\boldsymbol{\Delta}s^y.
\end{array}
\end{equation}
Note that $\mathcal{A}_k$ is a purely real matrix at $k=0,\pi$. Hence, the phase $\theta_k$, defined by $\exp(i\theta_k)=\det(\mathcal{A}_k)/|\det(\mathcal{A}_k)|$ can only be an integer multiple of $\pi$ at these momenta. This means that the following winding number
\begin{equation}
Q\equiv\frac{1}{\pi}\int_{k=0}^{k=\pi}d\theta_k
\end{equation}
is an integer topological invariant, as it can only change when $\det(\mathcal{A}_k)$ goes through zero, i.e. when the gap closes.

\subsection{Differential Conductance}
We wish to calculate the differential conductance into the system from a lead coupled to one end of the wire. We describe the system using the Hamiltonian $H^{\scriptscriptstyle{\rm OBC}}$, which is the same as the Hamiltonian $H^{\scriptscriptstyle{\rm HF}}$ from Eq.~\eqref{eq:HF_Hamiltonian}, but written using real space basis, and with open boundary conditions. We define the $8L\times8L$ BdG Hamiltonian $\mathcal{H}^{\scriptscriptstyle{\rm OBC}}$ such that
\begin{equation}
\begin{array}{ccc}
H^{\scriptscriptstyle{\rm OBC}}=\displaystyle\sum_{m,n}\varphi^\dag_m\mathcal{H}^{\scriptscriptstyle{\rm OBC}}_{mn}\varphi_n &,& \varphi^\dag=\begin{pmatrix}\psi^\dag&\psi\end{pmatrix}
\end{array},
\end{equation}
where $\psi^\dag_{m=4i+2s+\sigma}\equiv c^\dag_{\sigma,i,s}$ creates a particle with spin $s$ on site $i$ of chain $\sigma$. We couple the system to a lead with a single spin-degenerate mode, with both particle and hole degrees of freedom, such that the scattering matrix is a $4\times4$ unitary matrix, and is given by~\cite{datta1997electronic}
\begin{equation}
S(E)=\mathds{1}-2\pi iW^\dag\left[E-\mathcal{H}^{\scriptscriptstyle{\rm OBC}}+i\pi WW^\dag\right]^{-1}W ,
\end{equation}
where $W$ is an $8L\times4$ matrix describing the coupling of the lead to the system. In our case we choose the lead to be coupled only to the first site of chain $a$. The differential conductance is given by~\cite{shelankov1980resistance,blonder1982transition}
\begin{equation}
G(E)=\frac{e^2}{h}\left[2-\Tr(r^\dag_{ee}r^{}_{ee})+\Tr(r^\dag_{eh}r^{}_{eh})\right],
\end{equation}
where
\begin{equation}
S(E)\equiv \begin{pmatrix}r_{ee}&r^{}_{eh}\\ r_{he}&r^{}_{hh}\end{pmatrix}.
\end{equation}
Here, $r_{ee}$ and $r_{eh}$ are $2\times2$ matrices in spin-space, describing normal reflection and Andreev reflection respectively. We note that since single electrons cannot be transmitted into the bulk superconductor, and since there is only one lead, one necessarily has $\Tr(r^\dag_{ee}r^{}_{ee})+\Tr(r^\dag_{eh}r^{}_{eh})=2$.

\bigskip

\end{widetext}
\end{document}